\documentclass[%
 aip,
 preprint,%
]{revtex4-1}

\usepackage{amsmath}
\usepackage{amsfonts}
\usepackage{amssymb}
\usepackage{bm}
\def\*#1{\mathbf{#1}}
\usepackage[usenames,dvipsnames]{color} 
\usepackage{ulem}
\usepackage{graphicx}
\usepackage{epstopdf}

\usepackage[utf8]{inputenc}
\DeclareUnicodeCharacter{2010}{-}% support older LaTeX version
\DeclareUnicodeCharacter{2212}{-}

\usepackage{chemformula}
\usepackage{csquotes}
\usepackage[sort&compress]{natbib}
\usepackage{ulem}
\usepackage [english]{babel}
\usepackage[utf8]{inputenc}
\usepackage[T1]{fontenc}

\usepackage{gensymb}
\usepackage{lineno}

\usepackage{hyperref}
\hypersetup{
  colorlinks   = true, %Colours links instead of ugly boxes
  urlcolor     = black, %Colour for external hyperlinks
  linkcolor    = red, %Colour of internal links
  citecolor   = blue %Colour of citations
}

\begin{document}

\title{Structure and vibrational properties of sodium silicate glass surfaces}

\author{Zhen Zhang}
\author{Simona Ispas}
%\email[Corresponding author: ]{simona.ispas@umontpellier.fr}
\affiliation{ 
 Laboratoire Charles Coulomb, University of Montpellier, CNRS, F-34095 Montpellier, France
}%

\author{Walter Kob}
\email[Corresponding author: ]{walter.kob@umontpellier.fr}
%% \homepage{http://www.Second.institution.edu/~Charlie.Author.}
\affiliation{ 
 Laboratoire Charles Coulomb, University of Montpellier, CNRS, F-34095 Montpellier, France
}
%\affiliation{
%Institut Universitaire de France, 75231 Paris Cedex 05, France
%}

\date{\today}

\begin{abstract} 
Using molecular dynamics simulations we investigate the dependence of the structural
and vibrational properties of the surfaces of sodo-silicate glasses 
on the sodium content as well as the nature of the surface. Two types
of glass surfaces are considered: A melt-formed surface (MS) in which a liquid
with a free surface has been cooled down into the glass phase and a
fracture surface (FS) obtained by tensile loading of a glass sample. We find that the MS is more abundant in Na and non-bridging oxygen atoms than the
FS and the bulk glass, whereas the FS has higher concentration
of structural defects such as two-membered rings and under-coordinated
Si than the MS. We associate these structural differences to the
production histories of the glasses and the mobility of the Na ions. It is also found that for
Na-poor systems the fluctuations in composition and local atomic
charge density decay with a power-law as a function of distance from the
surface while Na-rich systems show an exponential decay with a typical
decay length of $\approx2.3$~\AA. The vibrational density of states shows that the presence of the
surfaces leads to a decrease of the characteristic frequencies in the system. The 
two-membered rings give rise to a pronounce band at $\approx880$~cm$^{-1}$
which is in good agreement experimental observations.

\end{abstract}

%\pacs{64.70.kj,63.50.Lm,64.70.Q-}

\maketitle

\section{Introduction}
\label{sec_I}

Silicate glasses are not only ubiquitous in many
technical applications and in our daily life but also in the
focus of many scientific studies.~\cite{shelby_introduction_2005,binder_glassy_2011,varshneya2013fundamentals} %The vast majority of these investigations have considered bulk glasses while the properties of glass surfaces have obtained significantly less attention~
%WK:We need more references here, notably newer ones. 
%ZZ: we need to narrow down the range. 
For many
practical applications of glasses, such as displays of electronic devices and
biomedical containers, the properties of the glass surface are highly
important since the coating and structuring of the surface allows to
devise novel functional materials.\cite{hench1978physical,Pantano1989,bach1997advanced,bocko1991surface,dey2016cleaning,zheng2019protein} %~\cite{XXX}. 
A further motivation to study the surfaces of glasses is that
such investigation allow to obtain a better understanding of the
failure mechanisms of bulk glasses, since very often fracture starts
at the surface defects of the sample.~\cite{anderson_fracture_2017,meyers_mechanical_2008} 
%{\color{blue}WK:Book on fracture}. 
Analyzing the post-mortem fracture surface of broken glass allows
thus to gain insight on the origin of the failure and the
way the fracture front propagates, knowledge that are valuable
for a deeper understanding of fracture of amorphous materials.~\cite{ciccotti_stress-corrosion_2009,rountree_unified_2007,gupta_nanoscale_2000,sarlat_frozen_2006,bonamy_scaling_2006,ponson_two-dimensional_2006,wiederhorn_roughness_2007,guin_fracture_2004,bonamy_failure_2011,wiederhorn_griffith_2013} 
%{\color{blue}WK: Add more general ref. on the fracture of amorphous materials.} ZZ: It cannot be too genernal since there are 1000 papers out there.

One of the primary goals of surface characterization is
to determine the composition and microstructure of the
sample. In experiments this can be done, e.g., by spectroscopic techniques
such as low-energy ion scattering (LEIS) spectroscopy,
X-ray photoelectron spectroscopy (XPS), or atomic force microscopy
(AFM).~\cite{kelso_comparison_1983,almeida_low-energy_2014,almeida_low-energy_2016,cushman2018low,radlein1997atomic,poggemann2001direct,poggemann2003direct,frischat2004nanostructure} 
Ion scattering spectroscopy study have revealed that
the fracture surface of silica glass shows an abundance of oxygen
atoms, and that the fracture surface of potassium trisilicate glass
has a potassium concentration that is higher than the one of the bulk
composition,~\cite{kelso_comparison_1983} a result which was attributed
to the charge shielding on the surface. More recent studies using LEIS
investigated the melt-formed and fracture surfaces of binary silicate
glasses.~\cite{almeida_low-energy_2014,almeida_low-energy_2016} It was
shown that, when compared with the bulk composition, the melt-formed
surfaces are usually depleted of the modifier atoms (i.e., Na)
which was hypothesized to be a consequence of surface evaporation while the sample was still in the liquid state. In contrast to this, the fracture
surfaces were found to be enriched in alkali species, while depleted of
divalent barium (which was attributed the immobility of the Ba$^{2+}$ cations). The authors of that study also
investigated the depth profiles of elemental concentration and were
able to detect the presence of concentration gradient normal to the
glass surfaces.~\cite{almeida_low-energy_2016}
% {\color{blue}WK: That the gradient is normal to the surface is trivial. But what was the decay length etc?} ZZ: To my knowledge, not known.
Also spectroscopic studies are useful to
determine structural features of the glass surface. For example
infrared and Raman measurements allowed to identify an interesting
structural motif on the surfaces of silicate glasses, namely two-membered (2M)
rings,~\cite{morrow1976infrared,michalske1984slow,bunker1989infrared,dubois1993bonding,dubois1993reaction,grabbe1995strained}
i.e., closed loops of two oxygen and two silicon atoms. Such metastable rings,
which are absent in the bulk, are of particular interest since they form reactive sites on the surface.~\cite{michalske1984slow,bunker1989infrared,dubois1993reaction,dubois1993bonding,grabbe1995strained,rimola_silica_2013}
Also the AFM has proven to be a valuable tool for direct
imaging the structural features on glass surfaces with atomic
resolution, allowing to access structural information such as interatomic distances and grouping of atoms.~\cite{radlein1997atomic,poggemann2001direct,poggemann2003direct,frischat2004nanostructure} 
%{\color{blue} WK: Was this really atomic resolution or just nanometric resolution? Can we say something about the results they found?} 
%Such studies have, e.g., allowed to identify the geometric properties of the surface on larger scales, giving evidence that the fracture surface has fractal properties~\cite{XXX}.

To summarize, these experimental studies have given clear evidence that the
composition and structure of glass surfaces are different
from the ones of the bulk. In addition, the surface composition
and structure were also found to depend strongly on the processing
history. However, it should be noted that these spectroscopic data
are essentially semi-quantitative and the results depend also on the
environment under which the measurements were performed. As a consequence
we are at present still lacking a good understanding about the composition
and structure of the glass surfaces and how these properties depend on
the system considered.~\cite{ren2017bulk,mahadevan2020hydration}

In addition to experimental studies, computer simulations have been used to probe the
microscopic properties of glass surfaces, particularly for the case of
silica.~\cite{feuston1989topological,garofalini1990molecular,roder_structure_2001,rarivomanantsoa_classical_2001,du_molecular_2005,rimola_silica_2013}
To characterize the simulated surface one often defines a surface
layer, the thickness of which is usually determined from properties
such as the density profile in the orthogonal direction with respect to
the surface, see for example in Ref.~\citenum{roder_structure_2001}. These
simulation studies have revealed the presence of structural units
such as nonbridging oxygen, 2M rings and undercoordinated Si on the glass
surfaces, in qualitative agreement with the experimental findings. However,
the vibrational spectrum and other properties which require a reliable interaction potential 
%{\color{blue}WK: examples?} 
are rarely reported,~\cite{roder_structure_2001,ceresoli_two-membered_2000}
despite their relevance for experiments. Also, at present it remains
poorly understood how the change in glass composition, e.g., different
concentration of alkali oxides, affects the surface structure and other
related properties.~\cite{ren2017bulk,mahadevan2020hydration}

The objective of the present work is to investigate how the properties (structure, composition, vibrational spectra,...)
of the glass surface depend on the production history and the composition. To this end, we perform large scale 
atomistic simulations to produce sodium silicate glasses with varied content of Na$_2$O and compare the characteristics of the melt-formed surface (MS) with the ones of the fracture surface (FS) of the glasses.

The rest of the paper is organized as follows: In Sec.~\ref{sec_II}
we give the details of the simulations and the way we have defined and
analyzed the surface. Section~\ref{sec_III} is devoted to the obtained
results and related discussion while in Sec.~\ref{sec_IV} we summarize and draw conclusions of this work.

\section{Methodology}
\label{sec_II}

\subsection{Simulation details} \label{sec:sim}

\begin{figure}[!t]
\centering
\includegraphics[width=0.8\textwidth]{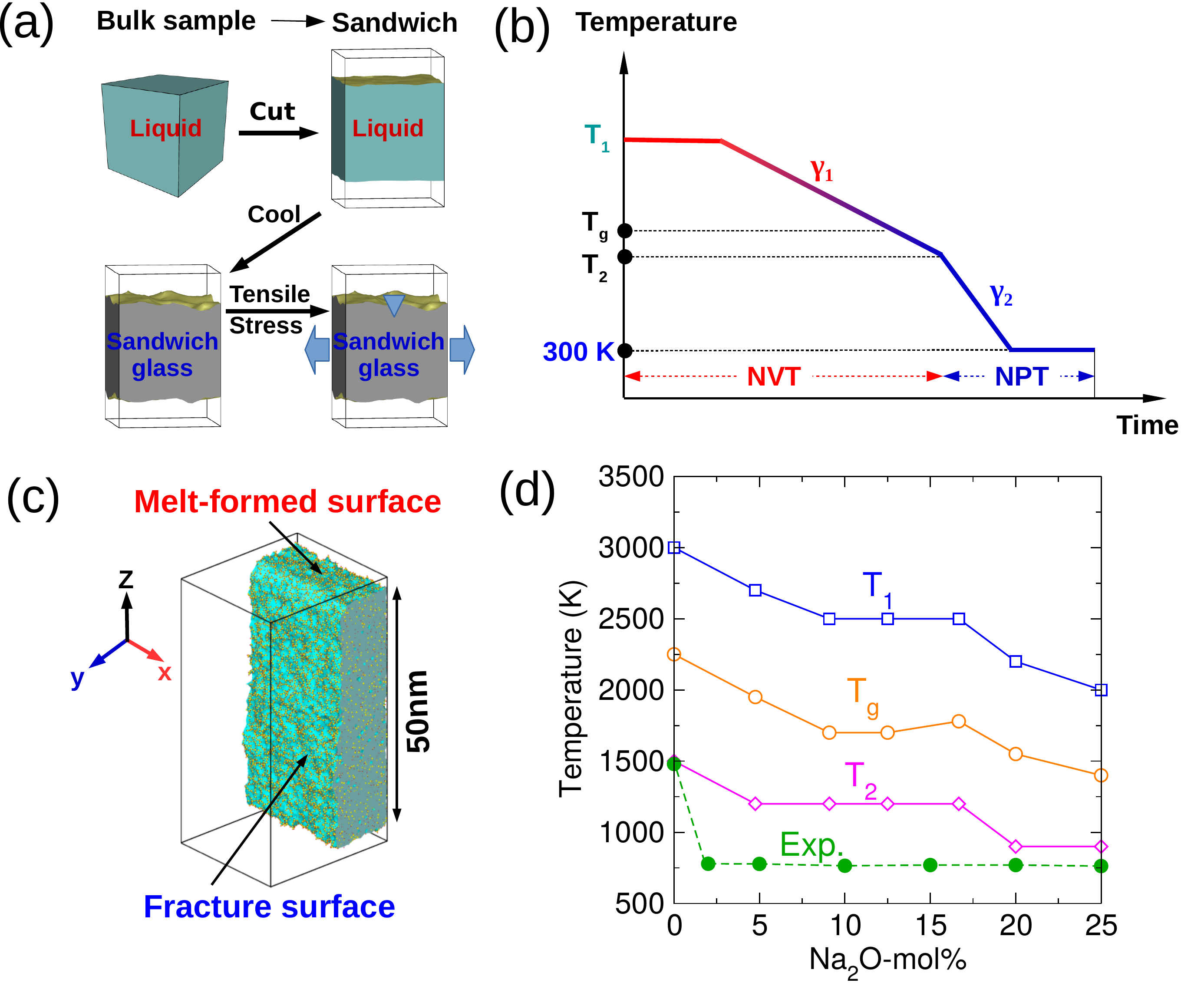}
\caption{Schematic drawing of the simulation procedures. (a) Preparation
of the sandwich glass sample. %{\color{blue}WK: Tensile => "Tensile stress"} 
(b) Temperature profiles of the melt-quench
procedure. See the text for the definitions of the various $T$s. (c)
A silica glass after fracture. The melt-formed and fracture surfaces
are indicated by the arrows. (d) Compositional dependence of various
characteristic temperatures in the simulations. The glass transition
temperature $T_g$ was determined as the $T$ where the extrapolated
total energy vs. temperature curves of the liquid and glass cross. The
uncertainty of the estimated $T_g$ is  about $\pm 50$~K. Also included in
the graph are the experimental $T_g$ (filled circles) measured by using
calorimetric method for Na$_2$O-xSiO$_2$~\cite{knoche_non-linear_1994}
and for SiO$_2$~\cite{richet_glass_1984} ($\Delta T_g = \pm 3 ^\circ$C). 
%{\color{blue}WK: connect all the experimental data by the green dashed line.}
}
\label{fig:cl protocols-glass-prod-frac}
\end{figure}

We have carried out molecular dynamics simulations to probe the
surface properties of SiO$_2$ and Na$_2$O-$x$SiO$_2$ (NS$x$) glasses. We choose the compositions with $x=3, 4, 5, 7, 10,$ and 20, and together with SiO$_2$ these compositions correspond to a Na$_2$O concentration that varies from 0 to 25 mole percent. % The chosen compositions correspond to a Na$_2$O concentration that varies from 0 to 25 mole percent, i.e.~$x=3, 4, 5, 7, 10,$ and 20. %{\color{blue}WK: correct?}. ZZ: yes
To start the simulations, we randomly placed around $2,300,000$ atoms in the simulation box which has a fixed volume
corresponding to the experimental value of glass density at room
temperature.~\cite{bansal_handbook_1986,fluegel_global_2007} [The glass density increases from 2.20~g/cm$^3$ for silica to 2.43~g/cm$^3$ for NS3 (25\% Na$_2$O).]
% {\color{blue}WK: Should we give the densities since this is an information that is not that easy to get if one doesn't have the book.} 
The dimensions of the
boxes were roughly $20$~nm$\times$30~nm$\times$50~nm. Such large
samples (surface areas are 600~nm$^2$ and 1000~nm$^2$ for the melt and
fracture surfaces, respectively) are necessary to obtain results with
high accuracy, in particular for the case of the fracture surfaces.
These samples, with periodic boundary conditions applied, were first melted and equilibrated at 6000~K for 80~ps
in the canonical ensemble ($NVT$) and then cooled and equilibrated at
a lower temperature $T_1$ (still in liquid state) for another 160~ps,
see Fig.~\ref{fig:cl protocols-glass-prod-frac}(d). The temperature $T_1$
ranges from 3000~K for SiO$_2$ to 2000~K for NS3 (25 mole\% Na$_2$O) and
its $x-$dependence reflects the fact that the viscosity of NS$x$ depends
strongly on $x$. Subsequently we cut the sample orthogonal to the $z-$axis,
and added an empty space, thus creating two free surfaces, i.e.~the sample had the geometry of
a sandwich, see~Fig.~\ref{fig:cl protocols-glass-prod-frac}(a).  Periodic boundary conditions were maintained in all three
directions. In order to ensure that the two free surfaces do not interact
with each other, the thickness of the vacuum layer varied from 6~nm
for silica to 14~nm for NS3. The samples with free surfaces were then
equilibrated at $T_1$ for 1.6~ns, a time span that is sufficiently long to allow
the reconstruction of the surfaces and the equilibration of the interior
of the samples. (The averaged displacement of Si in silica and NS3 is larger than 40~\AA\ and 6.2~\AA, respectively.)
%{\color{blue} WK: Can we say that the MSD is larger thanxx AA?} 
Following this equilibration the liquid samples were cooled
via a two-stage quenching: A cooling rate of $\gamma_1=0.125$~K/ps
was used to quench the sample from $T_1$ to a temperature $T_2$ and
a faster cooling rate $\gamma_2=0.375$~K/ps to cool it from $T_2$
to 300~K, see Fig.~\ref{fig:cl protocols-glass-prod-frac}(b). Finally,
the samples were annealed at 300~K for 800~ps. The temperature $T_2$
at which the cooling rate changes was chosen to be at least 200~K below
the simulation glass transition temperature $T_g$, see Fig.~\ref{fig:cl
protocols-glass-prod-frac}(d). At $T_2$, we also switched the simulation
ensemble from $NVT$ to $NPT$ (at zero pressure).

The described simulation protocol has the advantages that: 1) The fast cooling
below $T_g$ saves computer time while retaining statistically the same
structure as slow cooling; 2) The $NVT$ simulation in the high$-T$ range
helps to retain a regular shape of the sample. Below $T_g$ the sample
has more or less a regular shape and switching to the $NPT$ ensemble
allows to release internal stresses and facilitate local structural
rearrangements. In the following we will refer to the two surfaces of
the sandwich glass samples as the melt-formed surfaces (MS).

The sandwich glass samples were subsequently notched on one surface
and then subjected to an uniaxial strain in the $y-$direction until
complete fracture occurred, creating thus two fracture surfaces (FS),
see Fig.~\ref{fig:cl protocols-glass-prod-frac}(c). The cross section of
the introduced triangular notch had a width and height of 3~nm and 2~nm,
respectively (more details in Ref.~\citenum{zhang_thesis_2020}) and the
strain rate was chosen to be 0.5~ns$^{-1}$, a value that is small enough to obtain results that do not depend in significant manner on the rate.~\cite{zhang_thesis_2020,zhang_potential_2020}
The fracture simulations were done is the $NPT$ ensemble
at zero pressure, i.e., the pressures in the directions orthogonal to
the loading direction are set to zero, allowing the sample to relax in
the $x$ and $z$ direction. (The constant pressure ensemble is more close to real experiments than the constant volume ensemble since the latter induces artifact brittle fracture behavior of the sample under tension.~\cite{zhang_potential_2020,pedone_molecular_2008}) We note that the current simulation setup mimics
the plane stress condition, i.e., a thin slab in the $x-y$ plane with
the stress component $\sigma_{\alpha z}=0$, for $\alpha=x,y,z$.

For the simulations we used a pairwise effective potential named
SHIK which has been demonstrated to give a reliable description
of the structural and mechanical properties of sodium silicate
glasses.~\cite{sundararaman_new_2018,zhang_potential_2020} Its functional
form is given by

\begin{equation}
V(r_{ij}) =  \frac{q_iq_je^2}{4\pi \epsilon_0 r_{ij}} +
A_{ij}e^{-r_{ij}/B_{ij}} - \frac{C_{ij}}{r_{ij}^6} \quad ,
\label{eq:potential}
\end{equation}

\noindent
where $r_{ij}$ is the distance between two atoms of species $i$ and $j$.
This potential uses partial charges $q_i$ for different atomic species:
The charges for Si and Na are, respectively, fixed at 1.7755$e$ and
0.5497$e$, while the charge of O depends on composition and is given by
ensuring charge neutrality of the sample, i.e.,

\begin{equation} 
q_{\rm O}=\frac{(1-y) q_{\rm Si}+2y
q_{\rm Na}}{2-y}, 
\end{equation} 

\noindent
where $y$ is the molar concentration of Na$_2$O, i.e., $y=(1+x)^{-1}$. The other parameters
of the potential, $A_{ij}$, $B_{ij}$ and $C_{ij}$, occurring in
Eq.~(\ref{eq:potential}) are given in Ref.~\citenum{zhang_potential_2020}. It
is also worth to mention that these parameters of the SHIK potential
were optimized by using bulk properties obtained from experiments and
\textit{ab initio} calculations.~\cite{sundararaman_new_2018} To the
best of our knowledge it is the first time that this potential is used
to study dry surfaces of glasses.

Temperature and pressure were
controlled using a Nos\'e-Hoover thermostat and
barostat.~\cite{nose_unified_1984,hoover_canonical_1985,hoover_constant-pressure_1986}
All simulations were carried out using the Large-scale Atomic/Molecular Massively Parallel Simulator software~\cite{plimpton_fast_1995} with a time step of 1.6~fs. The results
presented in the following sections correspond to one melt-quench sample
for each composition. However, we emphasize that the system sizes considered in this study are sufficiently large to make sample-to-sample fluctuations negligible. For the MS, the results for the two surfaces on the top and bottom
sides of the glass sample were averaged. For the FS, four surfaces,
resulting from two independent fracture (by changing the location of
the notch), were averaged. For each sample the number of surface atoms was 
typically around 11000 for the MS and 18000 for the FS.% {\color{blue}WK: What is XXX?}

\subsection{Construction of the geometric surface} 
\label{sec:geo-surf} 

\begin{figure}[ht]
\includegraphics[width=1.0\textwidth]{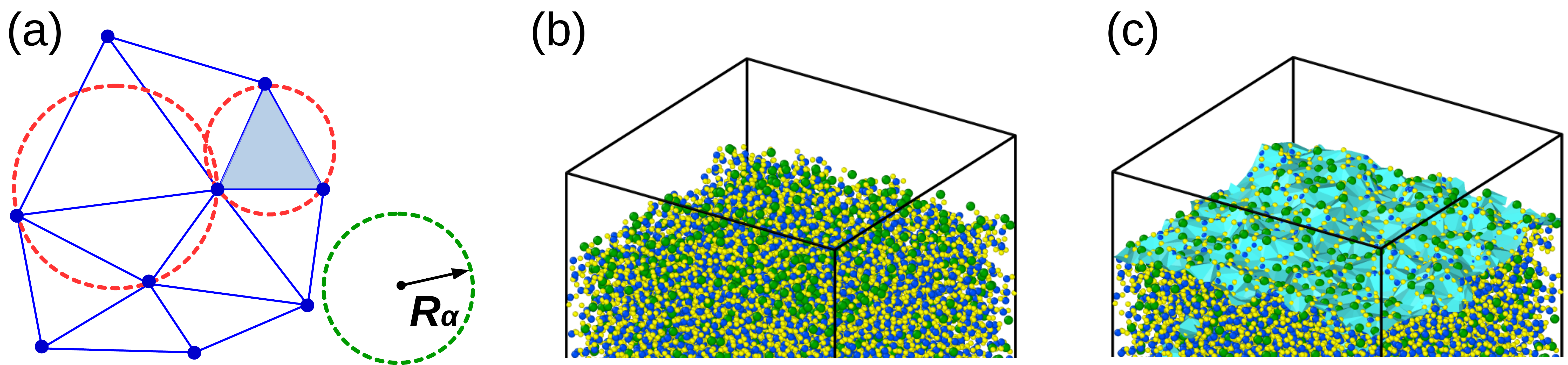}
\caption{Schematics of the procedure for constructing the geometric surface. 
(a) Delaunay triangulation of a set of points. The reference probe sphere
(with radius $R_\alpha$) is also shown. The triangular element whose
circumsphere is smaller than $R_\alpha$ is classified as solid (e.g.,
the shaded triangle). (b) Atomic model of the sandwich sample at 300
K. (c) Constructed polyhedral surface mesh. The shown sample is NS10
and the constructed surface is a melt-formed surface.
%{\color{blue}WK: I propose to remove the text above the panels and to leave only a,b,c.}
}
\label{fig:cl schematic-geometric-surf}
\end{figure}

In atomistic simulations, constructing the surface of a solid
corresponds to define the geometric boundary of a set of points
in space (the atoms) which allows to divide volume into solid
and empty regions. Figure~\ref{fig:cl schematic-geometric-surf}
shows schematically the procedure for constructing the geometric
surfaces. To start we have used the alpha-shape method proposed
by Edelsbrunner and M\"ucke~\cite{edelsbrunner_three-dimensional_1994}
to construct a surface mesh. For the two-dimensional case, this method
relies on the Delaunay triangulation (DT) of the input point set, see
Fig.~\ref{fig:cl schematic-geometric-surf}(a). For a given set $\bf{P}$
of discrete points, the triangulation DT($\bf{P}$) is done in such a way
that no point in $\bf{P}$ is inside the circumcircle of any triangle
in DT($\bf{P}$). For the three-dimensional case, the circumcircle
extends naturally to a circumscribed sphere which touches each of the
tetrahedron's vertices (Delaunay tetrahedrization).  All tessellation
elements are then tested by comparing their circumspheres to a reference
probe sphere, which has a radius of $R_\alpha$. The elements (with
circumsphere radius $R$) which satisfy $R < R_\alpha$ are classified as
solid, and the union of all solid Delaunay elements defines the geometric
shape of the atomistic solid. The mesh points which define this geometric
surface are the atoms on the outermost surface layer, 
Fig.~\ref{fig:cl schematic-geometric-surf}(c).

It is important to mention that the probe sphere radius $R_\alpha$ is the
length scale which determines how many details and small features of the
solid's geometric shape are resolved. Usually, the value of $R_\alpha$
should be chosen to represent the average spacing between the basic
structural units in a material.~\cite{stukowski_computational_2014} For
the investigated silicate glasses, we have chosen $R_\alpha=3.2$~\AA\
which corresponds approximately to the nearest neighbor distance between
two [SiO$_4$] tetrahedra. By visual inspection of the constructed surface,
we find that the chosen $R_\alpha$ allows to resolve fine surface features
while avoiding artificial holes in the constructed surfaces. We note,
however, that a small change of $R_\alpha$ (e.g. $\pm0.5$~\AA) will not
alter significantly the results presented in the following (more detailed
tests can be found in Ref.~\citenum{zhang_thesis_2020}). Visualization
of the atomic models and the surface mesh were realized by using the
OVITO software.~\cite{stukowski_visualization_2010} Finally we mention
that the procedure for constructing the FS is the same as the one for the MS. However, for the FS we have eliminated the atoms that
were closer than $\approx5$~nm to the top/bottom MS in order to avoid
the influence of these surfaces onto the properties of the FS.

\section{Results and discussion}
\label{sec_III}

\subsection{Composition of the vapor during the melt-quench process} 
\label{sec: vapor}

Before we discuss the properties of the glass surfaces it is instructive
to look at the simulation samples at high temperatures, notably at the
vapor phase, since this allows to understand better how the melt surfaces
are formed.

\begin{figure*}[tp]
\centering
\includegraphics[width=0.325\textwidth]{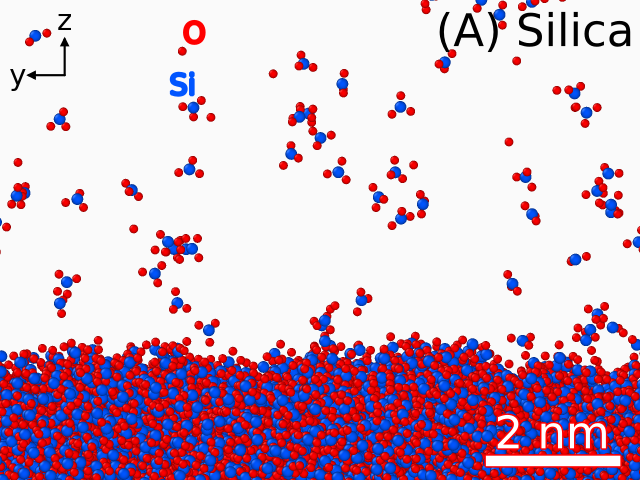}
\includegraphics[width=0.325\textwidth]{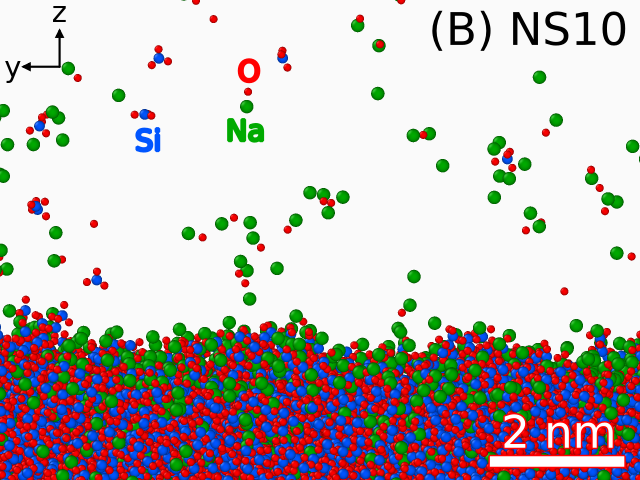}
\includegraphics[width=0.325\textwidth]{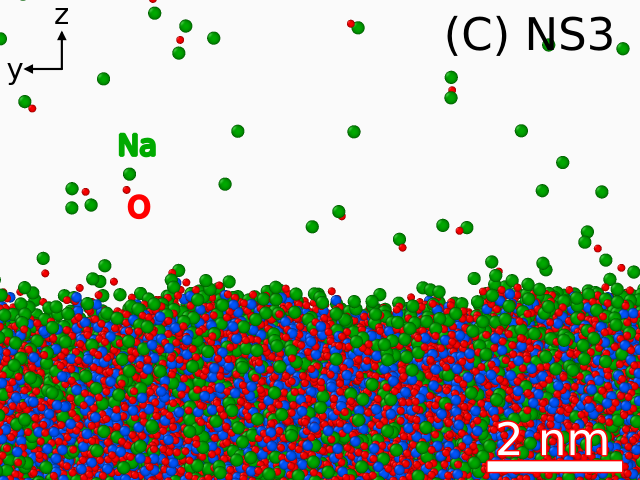}
\caption{Snapshots of the atomic structure near the surfaces of the 
liquid samples. (a) Silica at 3000~K. (b) NS10 at 2500~K. (c) NS3 at 2000~K.}
\label{fig:cl nsx-vapor-snapshots}
\end{figure*}

Figure~\ref{fig:cl nsx-vapor-snapshots} shows snapshots of a part of
the samples close to one of the surfaces at the temperature $T_1$,
i.e.~when the samples are still in the liquid state, see
Fig.~\ref{fig:cl protocols-glass-prod-frac}(d). For silica, panel (a),
one recognizes that most Si atoms in the vapor are surrounded by three O
atom and therefore one can expect that in the vapor the ratio between the
fractions of O and Si is around 3. With the addition of Na, NS10 panel
(b), the concentration of Si atoms in the vapor phase decreases quickly
since they are replaced to a large extent by the Na atoms which are
more volatile and still allow for (partial) charge compensation. The
trend of Si being replaced by Na continues if the Na content
is increased further, NS3 panel (c). One sees that for NS3 there are basically no Si atoms in the vapor phase.
This result agrees with the experimental finding that O and alkali (Na and
K) atoms will evaporate from the surfaces of alkali trisilicate glass
at elevated temperatures, while Si atoms remain at the surface and in 
the bulk.~\cite{kelso_spectroscopic_1985} It is worth to mention that
since the temperature considered for the three snapshots are not the same,
the reduction of Si content in the vapor may partially be attributed to
the decreased temperature for the samples containing Na (see below).

\begin{figure*}[t]
\centering
\includegraphics[width=0.325\textwidth]{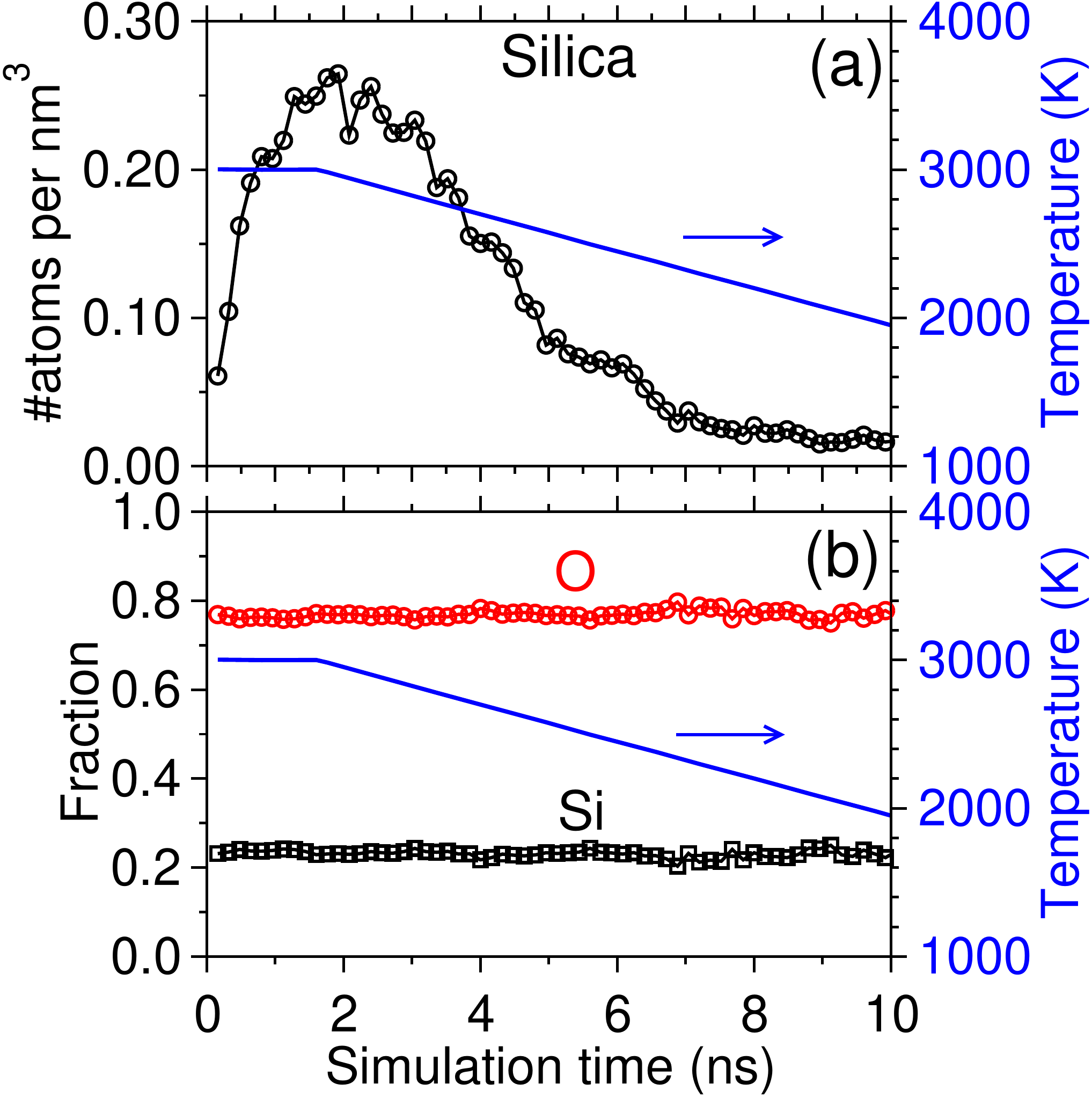}
\includegraphics[width=0.325\textwidth]{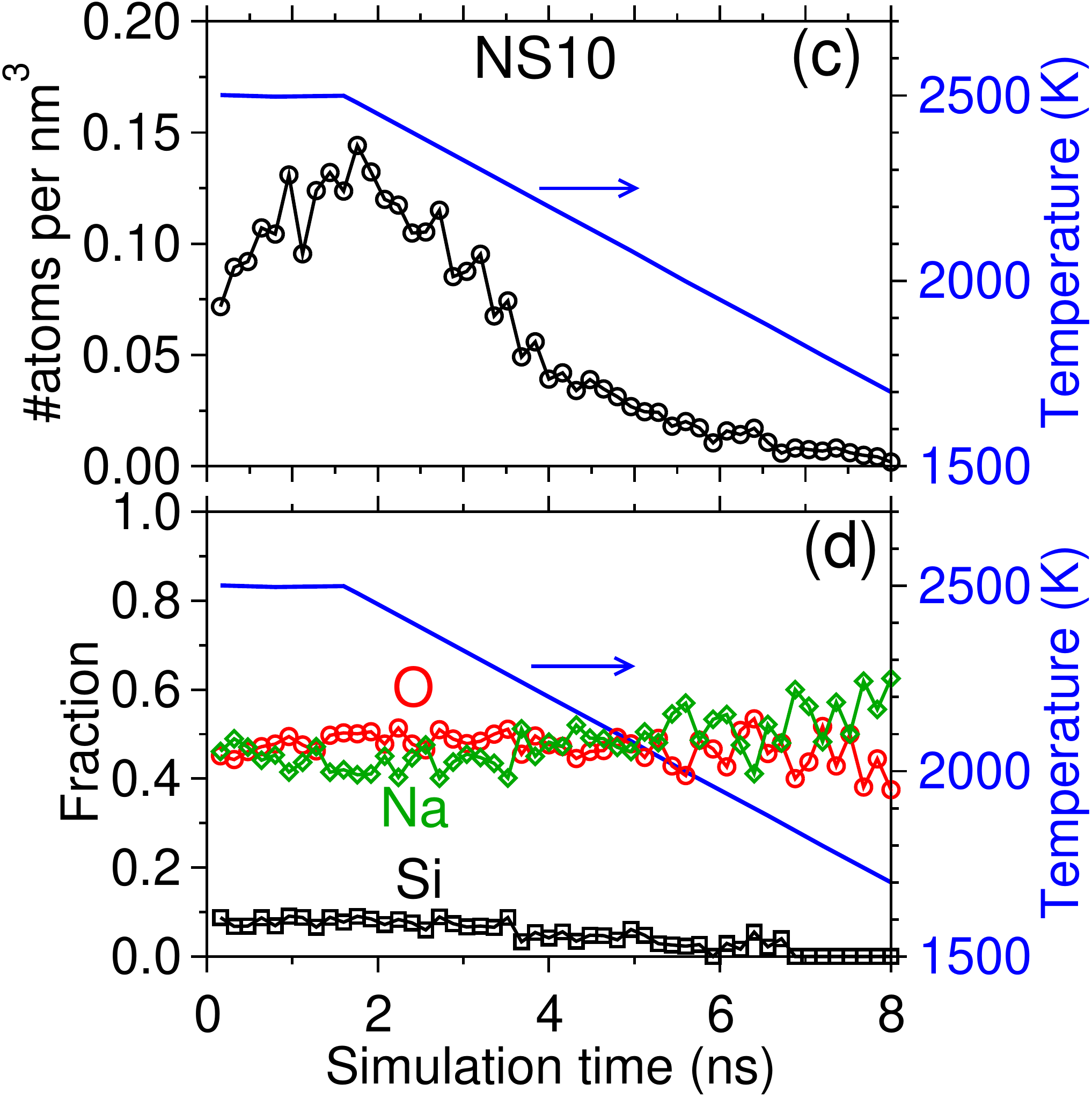}
\includegraphics[width=0.325\textwidth]{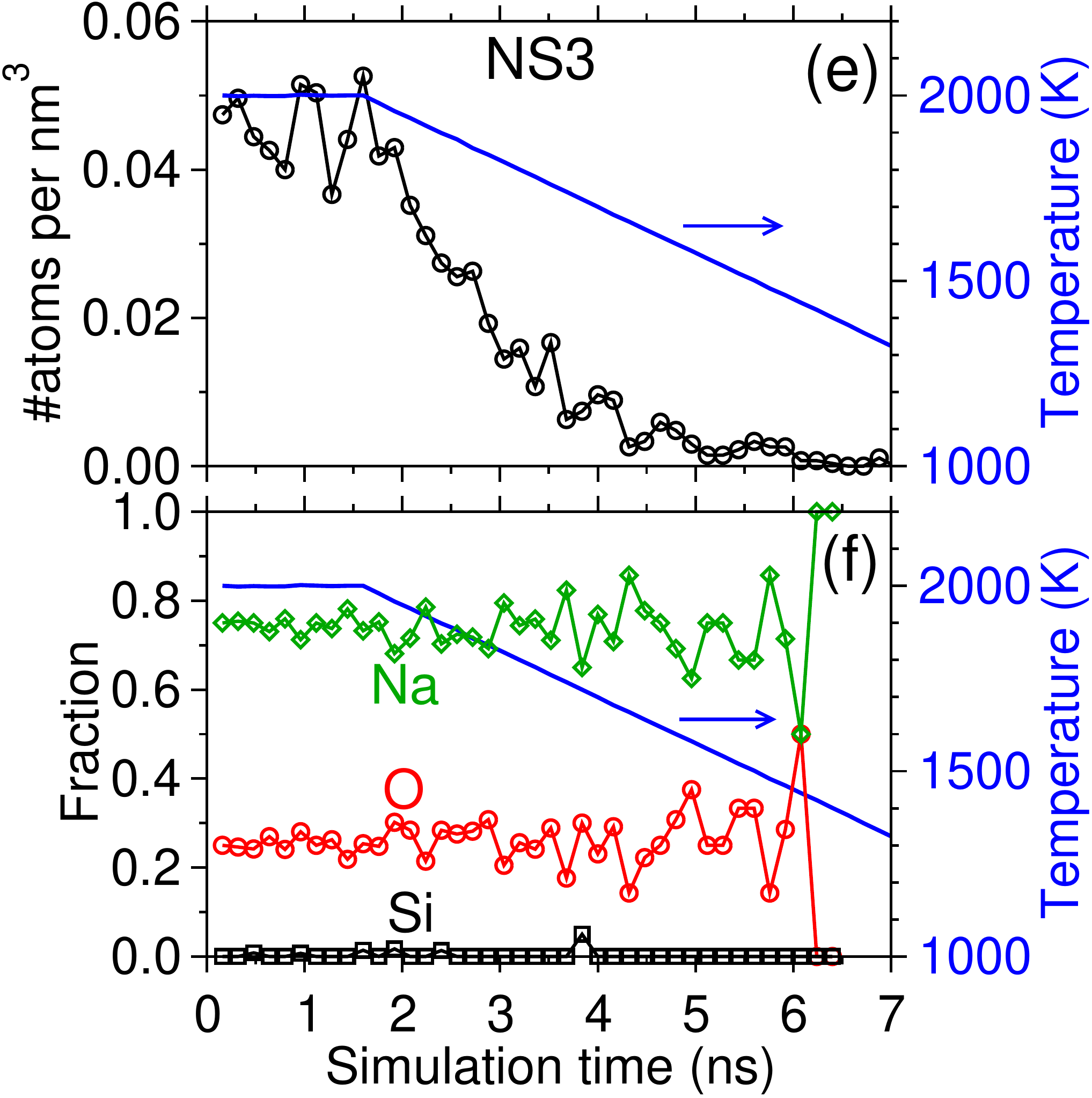}
\caption{Upper panels: Number density of atoms in the vapor of the samples during
the melt-quench process. Lower
panels: Fractions of different atomic species in the vapor. From left to
right the compositions are silica, NS10 and NS3, respectively.}
\label{fig:cl nsx-atom-num-frac-vapor}
\end{figure*}

In order to get a more quantitative understanding of these observations
we show in Fig.~\ref{fig:cl nsx-atom-num-frac-vapor} the number density
of atoms, $\rho_{\rm num}$, (upper panels) and the fractions of various
atomic species (lower panels) in the vapor phase as a function of time
during the equilibation (first 2~ns) and during the quench of the system. For silica, panel (a), one sees that $\rho_{\rm num}$ increases while
the system is at 3000~K, indicating that more and more atoms in the
near-surface region evaporate to the vacuum, demonstrating that the
sample is above the boiling point. Once the temperature is lowered, $\rho_{\rm
num}$ quickly decreases and for $T$ below 1400~K only very few atoms remain
in the gas phase. Interestingly one finds that the relative concentrations
of O and Si in the vapor are independent of temperature and that the ratio
between O and Si is close to 3, panel (b). This result indicates that,
at these temperatures, each Si atom moving in and out from the surface
is likely to be associated with three O atoms, instead of four as one
would expect for a [SiO$_4$] tetrahedral unit, in agreement
with the snapshot in Fig.~\ref{fig:cl nsx-vapor-snapshots}(a).

%Also for NS10, panel (c) one finds that at early times the number of atoms in the vapor phase increases but in comparison with SiO$_2$ the effect is less pronounced. For NS3 this evaporation is absent, showing that at the temperature considered the system is below its boiling point. Note that for NS3 the number density of atoms in the gas phase at, say, $T=2000$~K (panel (e)), is lower than the one found in NS10 (panel (c)), although the typical temperature scales such as $T_g$ are below the ones of NS10. This shows that the addition of sodium prevents the evaporation of the atoms from the surface, even if the dynamics speeds up. 
%{\color{blue}{WK: This reasoning is only correct if the empty space is the same. Is it?}} ZZ: No, we have already mentioned this in the simulation details. 
%{\color{red}{ZZ: I don't agree with the arguments in this paragraph. 1)for NS3, 2000~K is above the boiling point. otherwise you won't see atoms in the vapor. 2) on the contrary, I think Na helps the evaporation of atoms from the surfaces. Notice that there are differences in their equilibrating temperature and vacuum thickness. Please take a look at my original discussion about this result.}}
With increasing Na$_2$O content, panels (c) and (e), the density of atoms in the vapor decreases considerably relative to the silica one. This result is on one hand due to the reduced temperature of equilibration (Fig.~\ref{fig:cl protocols-glass-prod-frac}d) and on the other hand because of the increased volume of the vacuum layer.
Regarding the composition of the vapor one finds that for NS10, panel (d),  the
fraction of Si is around 10\% during the equilibration, and then decreases with lowering $T$, while
the Na fraction is around 50\% and this number increases slightly as the
temperature decreases. For NS3, panel (f), the enrichment
of Na in the vapor is even more pronounced than in NS10, reaching a
concentration of around 75\% (compared to the nominal fraction of Na
$=16.7\%$ for NS3), while the concentration of Si in the vapor becomes
negligible, a result that is consistent with the observation from the
snapshot in Fig.~\ref{fig:cl nsx-vapor-snapshots}(c). Interestingly,
one finds that the relative concentration between Na and O is around
3, independent of temperature.  %These results show that to a first approximation the concentration of the various species in the vapor phase is independent of temperature, i.e.~that the relative value of the partial pressures are independent of temperature. 

Finally we note that if one multiplies the concentration of the various atoms with their respective charges, one finds
that the total charge of the vapor phase is negative for silica and NS10, panels (b) and (d), while it is positive for NS3, panel (f). Consequently, the condensed phases are also not charge neutral. This charge non-neutrality of the two phases have consequences in the surface properties as we will see later.

%However, in practice this charge inbalance {\it per atom} is very small since the number of atoms in the vapor phase is much smaller than the one in the slab. Finally we point out that the temperatures at which $\rho_{\rm num}$ becomes (basically) zero is close the glass transition temperature of the sample, which are around 2200~K, 1700~K, and 1400~K for silica, NS10, and NS3, respectively. This shows that the temerature of condensation is not that different from $T_g$.

\subsection{Monolayer surface composition and structure}

Having understood the composition of the vapor phase at elevated
temperatures, we now focus on the surface properties at 300~K. To start
we characterize the local structure of the surfaces by the distributions
of interatomic distances and bond angles and compare these distributions
with the ones of the bulk glass. This bulk data is obtained by considering
only the atoms in a cube with side length of 120~\AA\ in the geometric
center of the sandwich glass, i.e.~these atoms have a distance
of at least 190~\AA\ from any free surface and thus can be expected
to reflect the bulk behavior of the glass sample.

 Figure~\ref{fig:cl
nsx-surf-dist-PDF-si} shows the distributions of nearest neighbor for
various pairs. For clarity, we show only the results for three representative compositions, namely silica, NS10 (Na-poor) and NS3 (Na-rich). For Si-O, panels (a-c), one recognizes that the distributions
for the surfaces and the bulk are very similar to each other, indicating
that the presence of the surface does not modify significantly the
bonding distance. Nevertheless one sees that the distribution for
the bulk is narrower than the one of the free surfaces and that the
FS has a distribution that is somewhat broader than the one of the
MS. The latter result is reasonable since the FS has more structural
defects (see below). For the Si-Si distances, panels (d-f), we find
for the surfaces a pronounced peak at $\approx2.4$~\AA\ , a feature
that is completely absent in the distribution for the bulk sample.
(Note that the MS curves for NS10 and NS3 are not shown since
they are too noisy as only a small number of Si atoms is found on the surface.)  This peak arises from the two-membered (2M)
rings in which two edge-sharing Si atoms (esSi) are connected by two
edge-sharing bridging O atoms (esBO), as already documented in previous
studies.~\cite{morrow1976infrared,michalske1984slow,bunker1989infrared,dubois1993bonding,dubois1993reaction,grabbe1995strained,ceresoli_two-membered_2000,roder_structure_2001,rarivomanantsoa_classical_2001,mischler2002classical,halbert2018modelling}%i.e.~a structural motif that is completely absent in the bulk. 
%From these graphs one also recognizes that the area under the 2M peak increase with increasing Na concentration which shows that with increasing Na these two-membered rings becomes the dominating local motif.

\begin{figure}[t]
\centering
\includegraphics[width=0.95\textwidth]{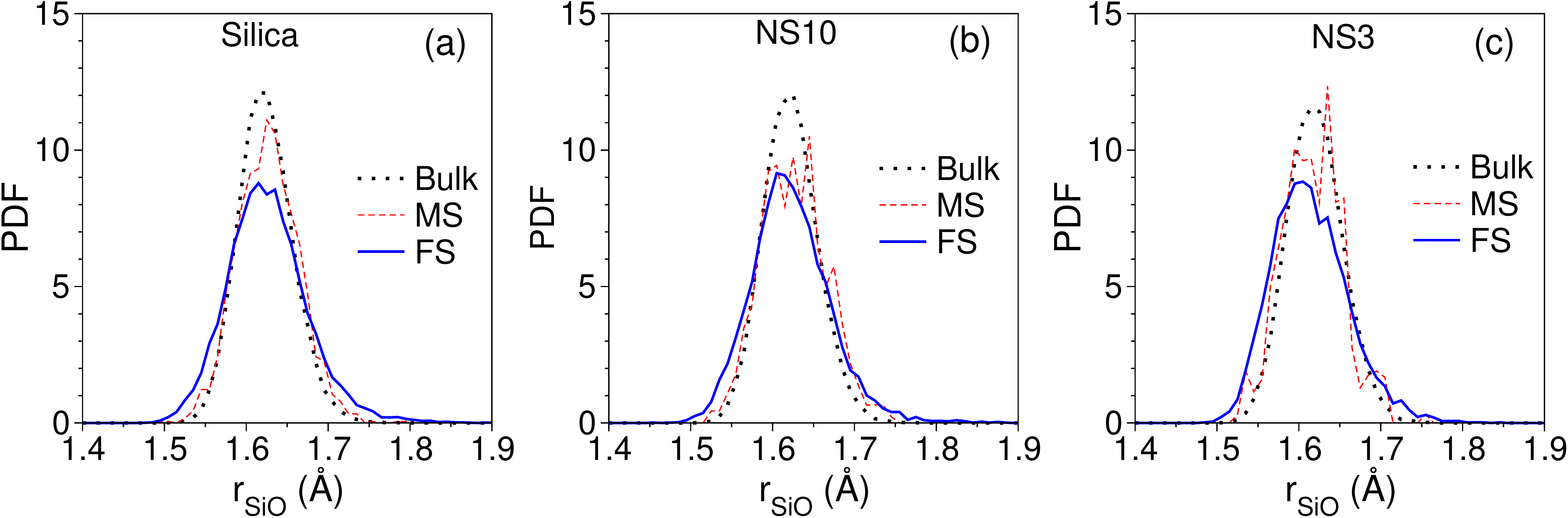}
\includegraphics[width=0.95\textwidth]{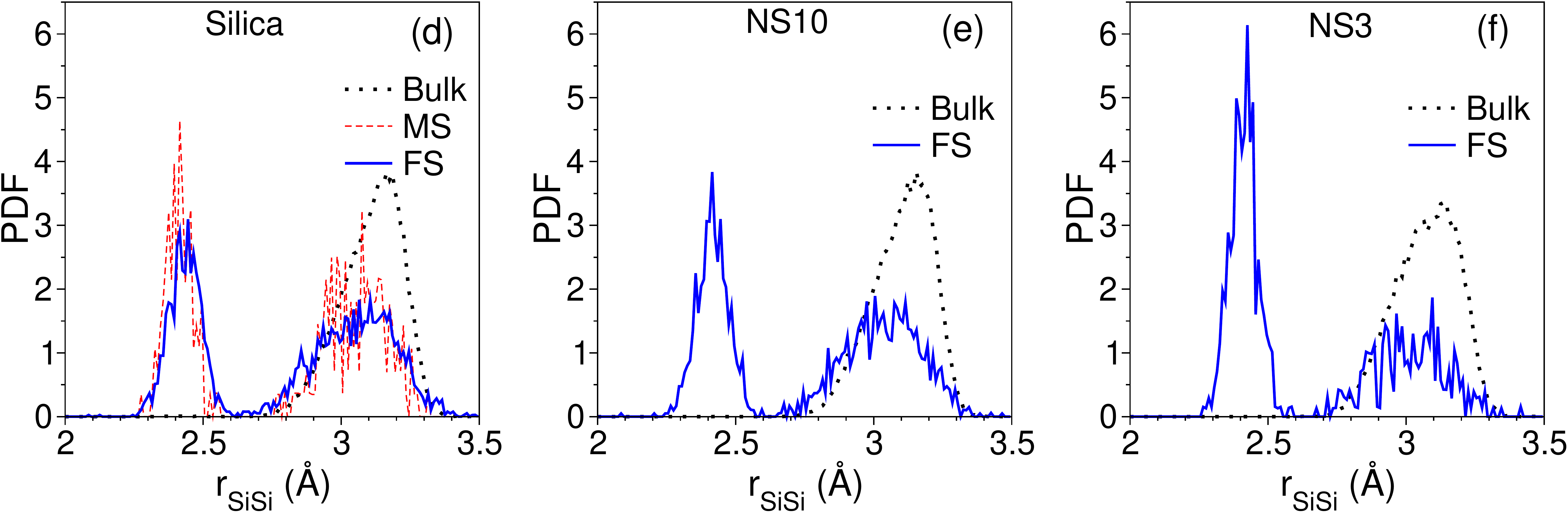}
\caption{ Probability distribution function (PDF) of nearest
neighbor distance. Top and bottom panels are for Si-O and Si-Si pairs,
respectively. From left to right, the compositions are silica, NS10, and
NS3, respectively.
% {\color{blue}{WK: I propose to label the y-axis as "PDF" and the x-axis as $r_{SiO}$ etc. Also, the bulk curve should be a bit bolder. It might be better to normalize the curves so that they become a probability (or are they already normalized?). Also in the lower row one has to update the (a)-(c) to (d)-(f).}}
}
\label{fig:cl nsx-surf-dist-PDF-si}
\end{figure}

\begin{figure}[t]
\centering
\includegraphics[width=0.8\textwidth]{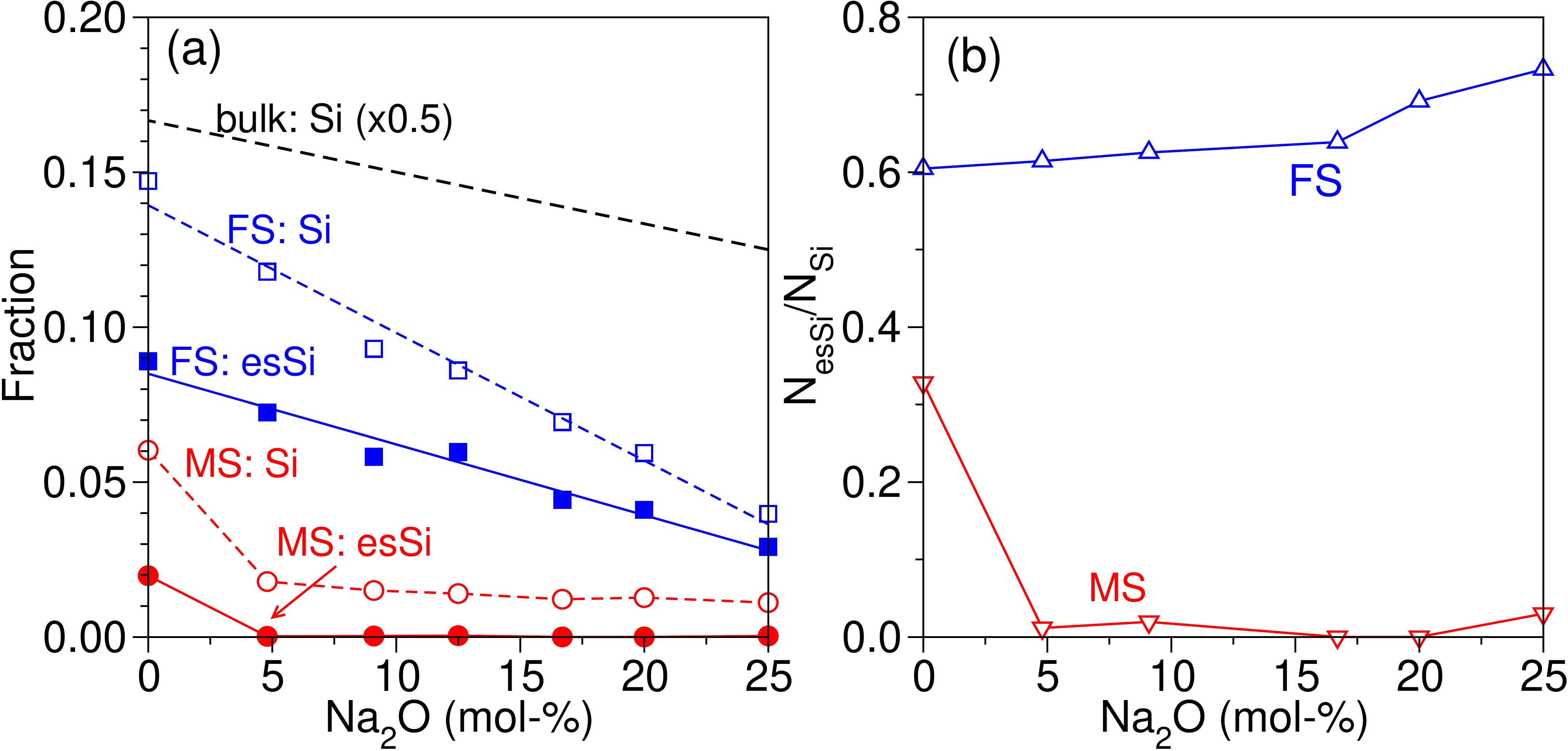}
\caption{(a) Fraction of Si and esSi on the surfaces. The bulk data are
multiplied by 0.5 to allow better comparison with the surface data. The
blue solid and dashed lines are linear fits to the FS data. (b) The ratio between the fractions of esSi and Si on the surfaces. }
\label{fig: nsx-surf-si-essi}
\end{figure}

Figure~\ref{fig: nsx-surf-si-essi}(a) shows the concentrations of Si and
esSi as a function of Na$_2$O content for the surfaces. As a reference we
have included in the graph also the concentration of Si in the bulk state
(black dashed line), which decreases linearly with the Na concentration.
The graph demonstrates that due to the enrichment of O and Na atoms on the
surfaces, the surface concentration of Si is considerably lower than the
bulk one. Furthermore one sees that for the FS the concentration of esSi
is considerably higher than the one of the MS, i.e.~the former surface
is more abundant in 2M-rings than the MS, a trend that can be understood directly
from the production process of the two types of surfaces. For the FS,
both the Si and esSi fractions decrease in a linear manner with Na$_2$O
content showing that on the surface the Si atoms are readily replaced
by Na atoms. 
The graph also demonstrates that the Si concentration in the MS depends very weakly on the Na$_2$O content. The
only exception is for the Na-poor compositions in that one observes drastic decrease in the concentration of Si at the surface if one goes from
pure silica to a glass with 5\% Na$_2$O. This result is
directly related to the fact that the MS is created in the liquid state
which allows more Na to diffuse to the surface and thus to reorganize
the surface structure by reducing the concentration of Si atoms and
hence the local stress.

The ratio between the fractions of esSi and Si on the two types of surface are plotted in Fig.~\ref{fig: nsx-surf-si-essi}(b). One recognizes that for the FS over 60\% of Si atoms are in the 2M-ring structures and this ratio slightly increases with increasing Na$_2$O content. The MS, by contrast, only has a negligible fraction of esSi on the surface, except for silica for which around 30\% of Si are esSi.
%Since the slope of the esSi data is smaller than the one of the Si curve, we conclude that at intermediate and high Na concentration most Si atoms are in 2M structures and at the highest Na concentration considered these motifs do become the most frequent ones (see also Fig.~\ref{fig:cl nsx-surf-strcut-compo}). 
We mention here that the FS was
generated by fast cracking ($\approx10^3$~m/s~\cite{zhang_thesis_2020})
at room temperature and therefore one can expect that only very little
reconstruction of the fracture surface has occurred after the crack
has passed. 
%{\color{blue}{WK: In principle the $T$ is not that low since close to the crack the temperature is high. But perhaps we don't need to mention this here.}}
%ZZ: two factors: temperature and time.  T may be ~1000K (i.e. still below Tg) but very short time.

\begin{figure}[th]
\centering
\includegraphics[width=0.65\textwidth]{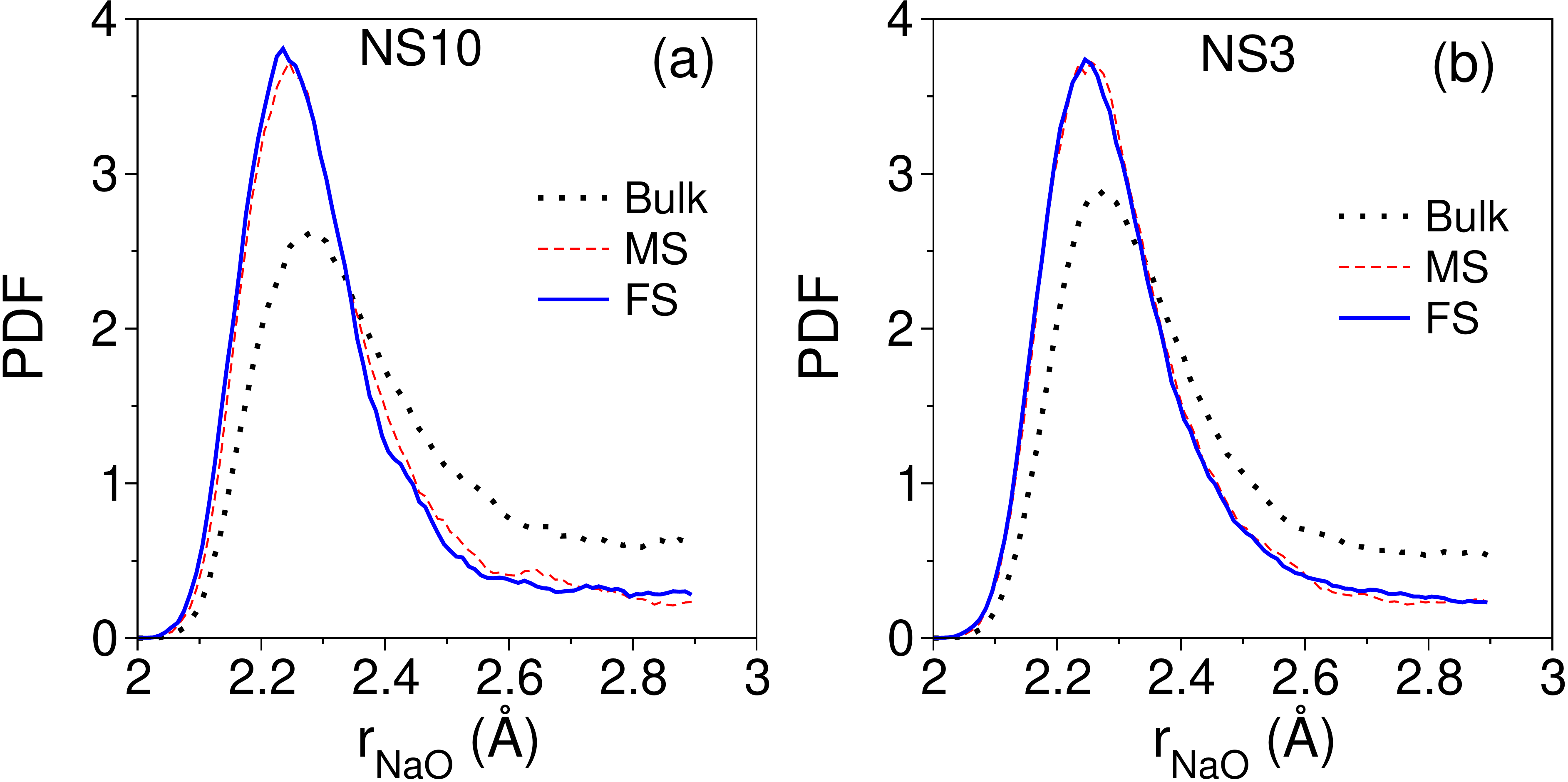}
\includegraphics[width=0.65\textwidth]{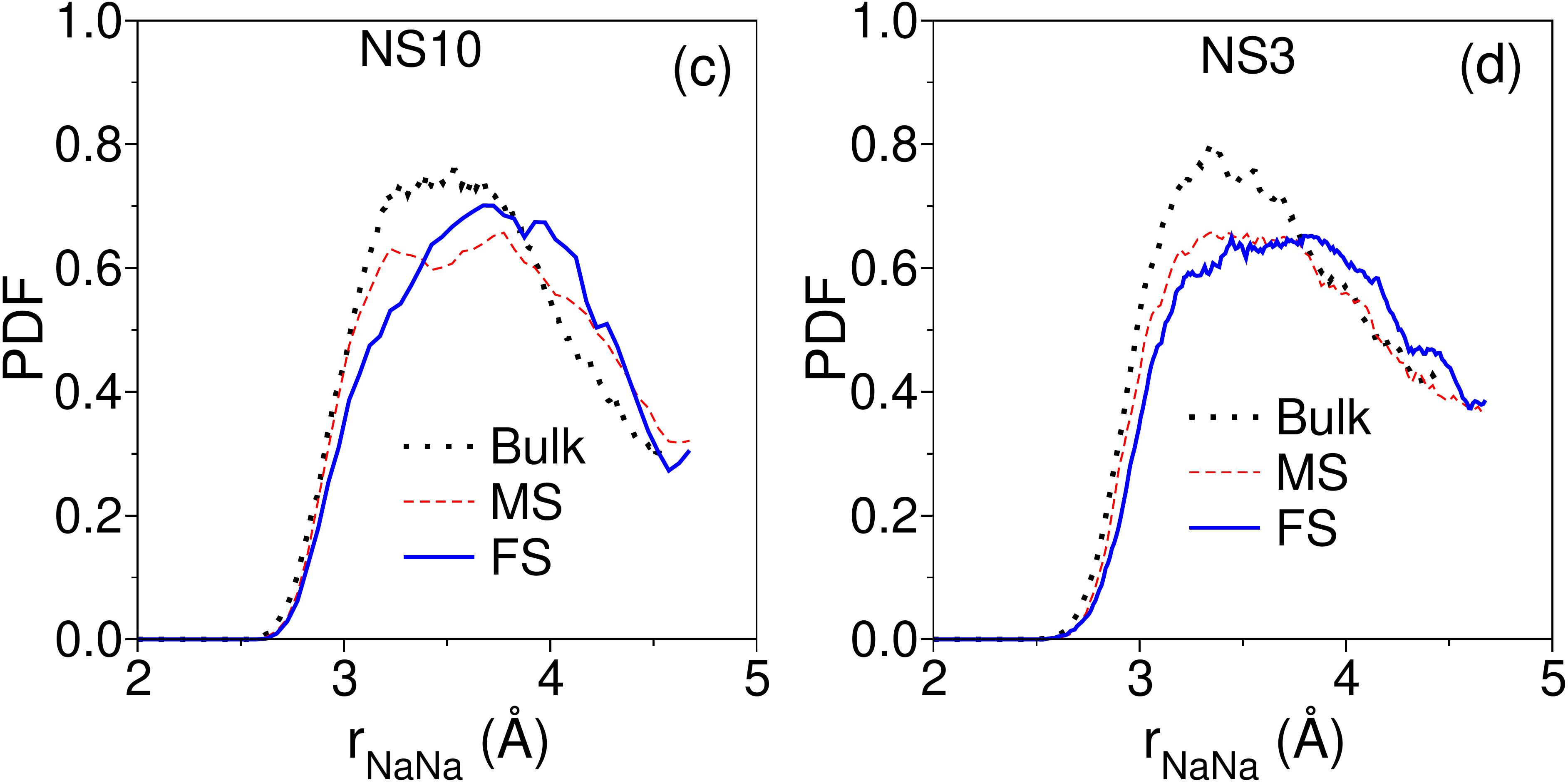}
\caption{ Probability distribution function (PDF) of nearest
neighbor distance. Top and bottom panels are for Na-O and Na-Na pairs,
respectively. Left and right columns are for NS10 and NS3, respectively. %{\color{blue}{WK: I propose to label the y-axis as "PDF" and the x-axis as $r_{NaO}$ etc. Also, the bulk curve should be a bit bolder. It might be better to normalize the curves so that they become a probability (or are they already normalized?).}}
}
\label{fig:cl nsx-surf-dist-PDF-na}
\end{figure}

\begin{figure*}[!t]
\centering
\includegraphics[width=0.45\textwidth]{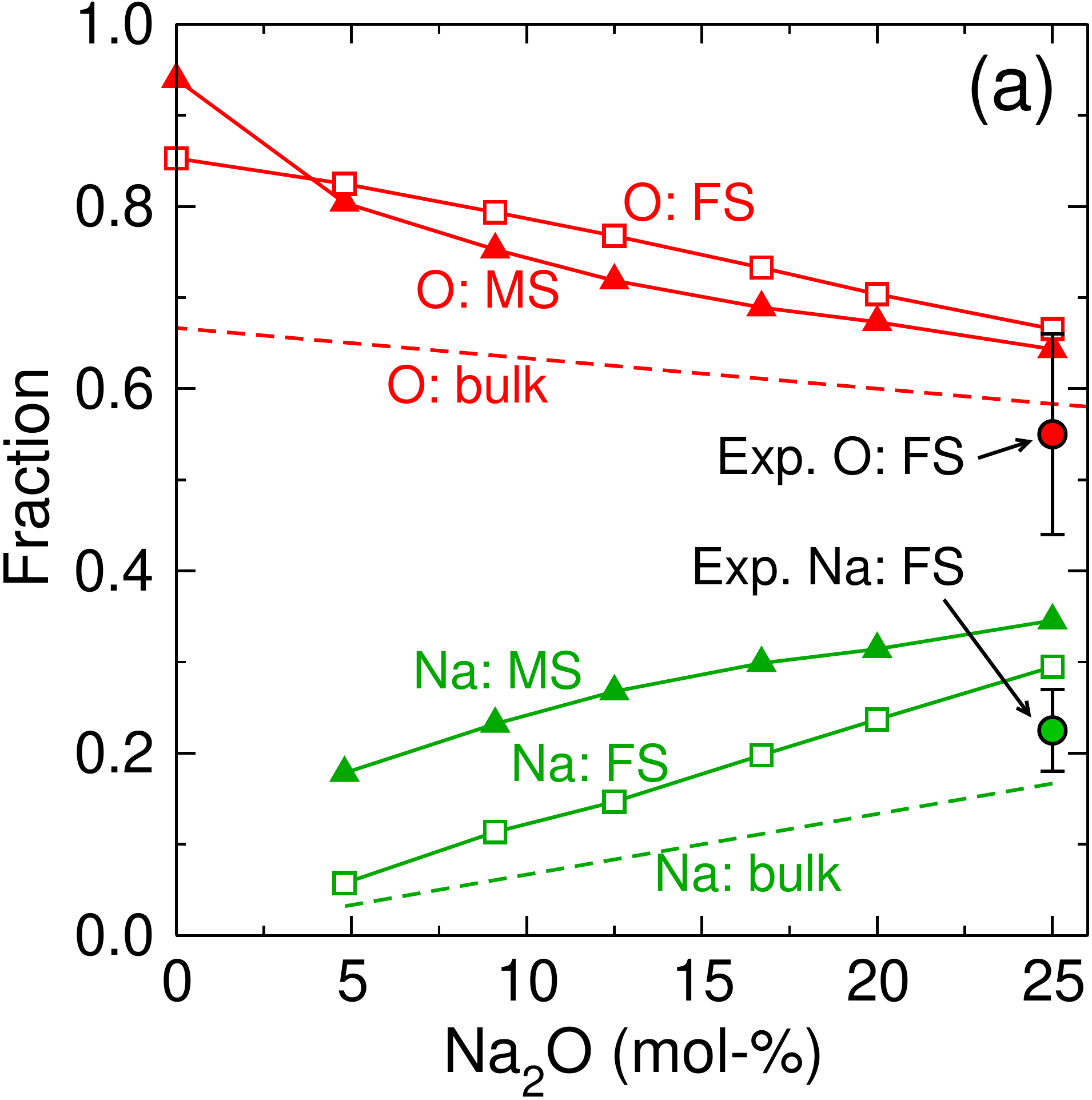}
\includegraphics[width=0.45\textwidth]{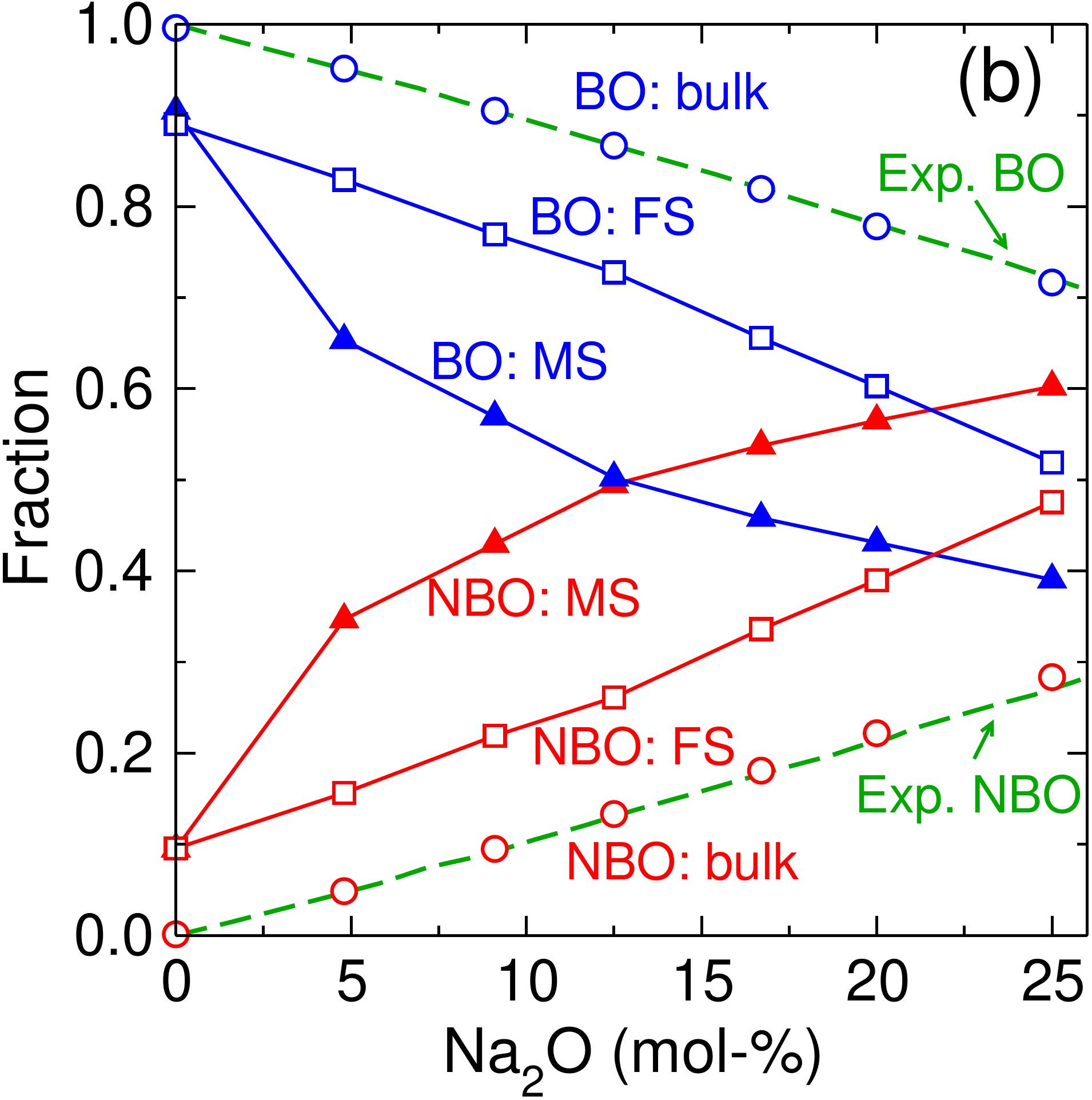}
\includegraphics[width=0.45\textwidth]{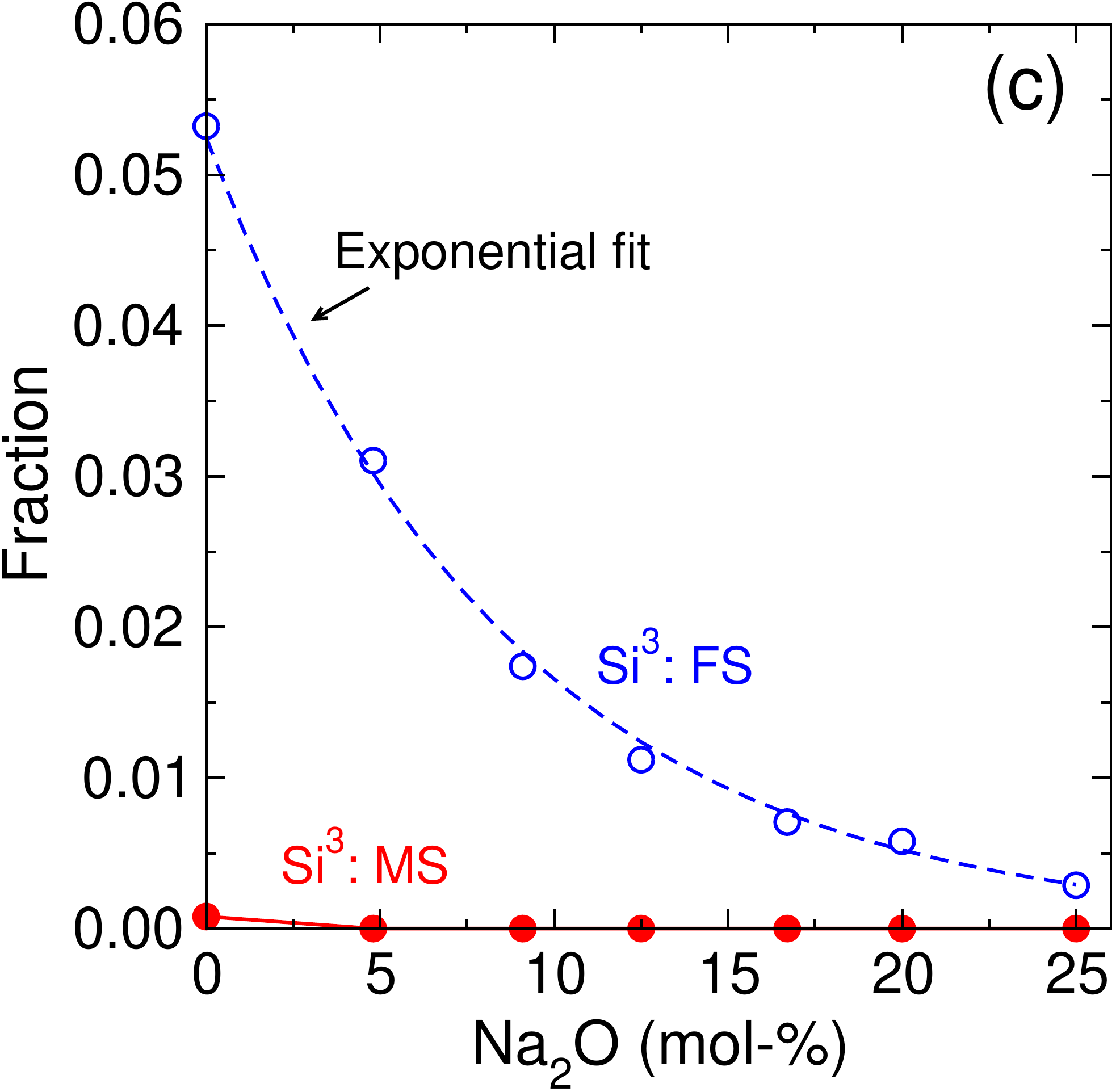}
\includegraphics[width=0.45\textwidth]{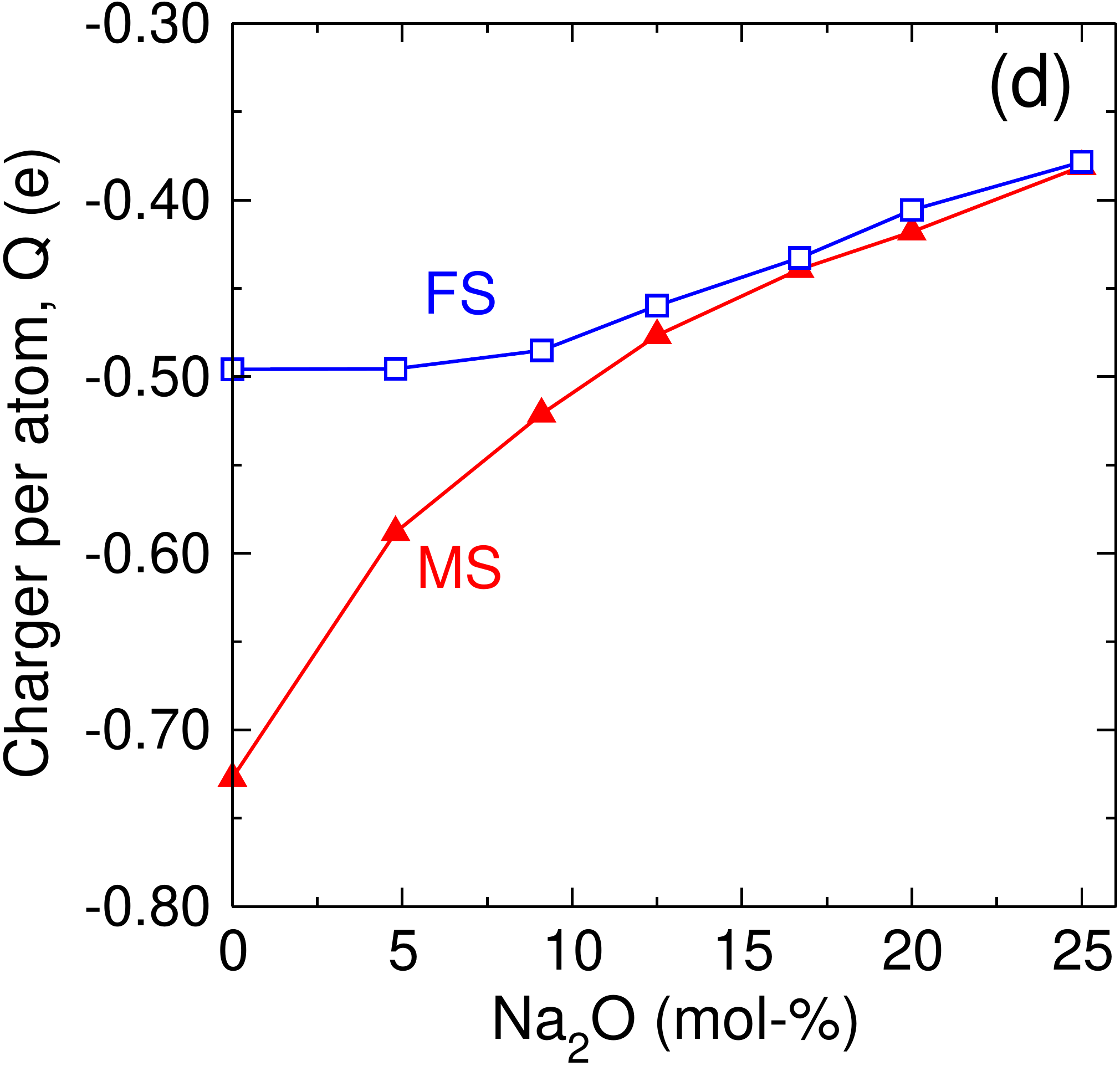}
\caption{Surface composition and structure. (a) Fraction
of O and Na atoms with respect to the total number of atoms on the surface. Experimental
data are taken from Ref.~\citenum{almeida_low-energy_2014}. (b) Fraction of BO and
NBO species with respect to the total number of O atoms on the surface. The
green dashed lines are fits to experimental data as give in
Refs.~\citenum{veal1982xps,maekawa1991structural,nesbitt_bridging_2011}. (c)
Fraction of undercoordinated Si defects with respect to the total number of Si atoms on the surface. The dashed line is an
exponential fit to the FS data. (d) Per-atom atomic charge on the surfaces. Error bars are standard deviation and
are smaller than the symbol size for all data points.
%{\color{blue}WK: In panel (a) we should remove the Na data point for concentration zero. In panel (b) the y-axis should read ``Charge per atom, Q (e)''.} ZZ: I would suggest keep them, because they help to show the pronounced compositional dependence at Na20<5%.
}
\label{fig:cl nsx-surf-strcut-compo}
\end{figure*}

Figure~\ref{fig:cl nsx-surf-dist-PDF-na} shows the distributions of the
nearest neighbor distances for the Na-O ($r_{\rm NaO}$) and Na-Na ($r_{\rm
NaNa}$) pairs. For Na-O, panels (a) and (b), one recognizes that the two
curves for the surfaces are nearly identical, indicating that the relative
arrangement of surface O and Na is independent of the nature of the
surface, despite the fact that for the two surfaces the Na concentration
is very different (see Fig.~\ref{fig:cl nsx-surf-strcut-compo}).
Furthermore we note that the distributions of the $r_{\rm NaO}$ distances
for the two surfaces are narrower and peak at a smaller distance than the one of the bulk. This
finding indicates that the most probable Na-O bond length on the surfaces
is slightly smaller than the typical value found in the bulk. The rationale is that on the surfaces the atoms have less constraints, and thus the Na and O atoms are more likely to form bonds that are energetically more favorable.

For the Na-Na distance, panels (c) and (d), one notices that also here
the distributions for the two surfaces are very similar but that now
these distributions are slightly broader than the one for the bulk. This
observation might indicate that on the surface the Na arrangement (with respect to the Si-O network) is
more disordered than in the bulk but it might also related to the increased local concentration of Na atoms.

\begin{figure*}[th]
\centering
\includegraphics[width=0.95\textwidth]{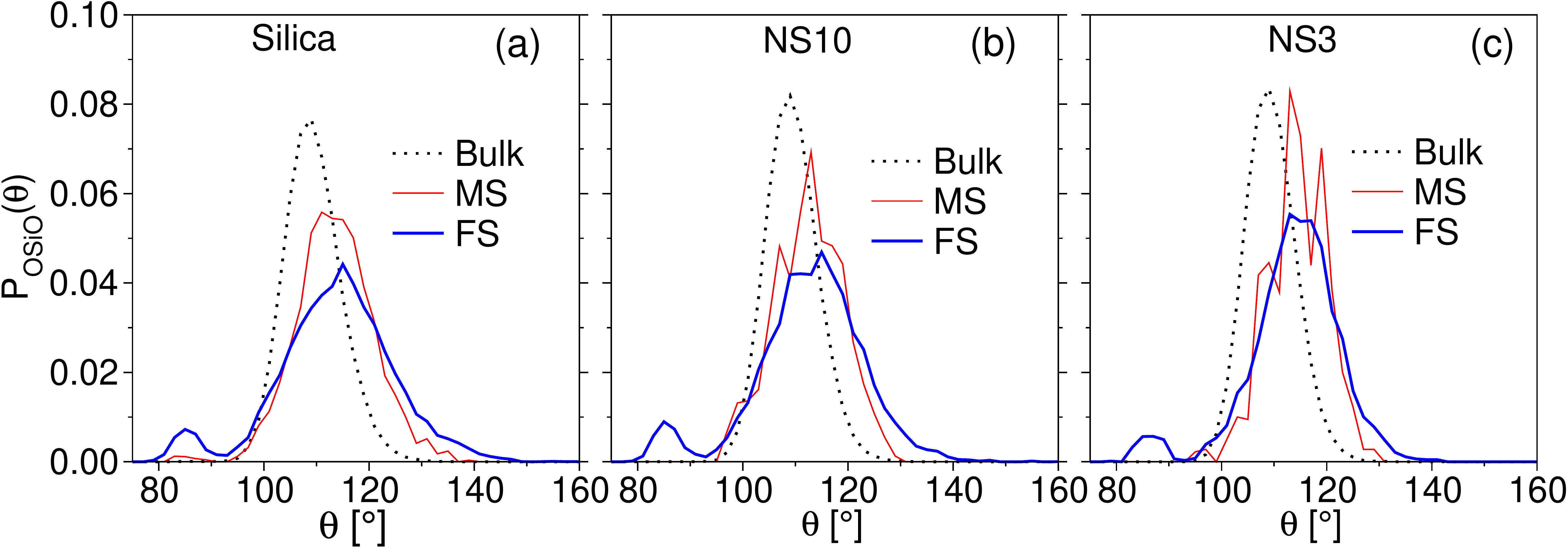}
\includegraphics[width=0.95\textwidth]{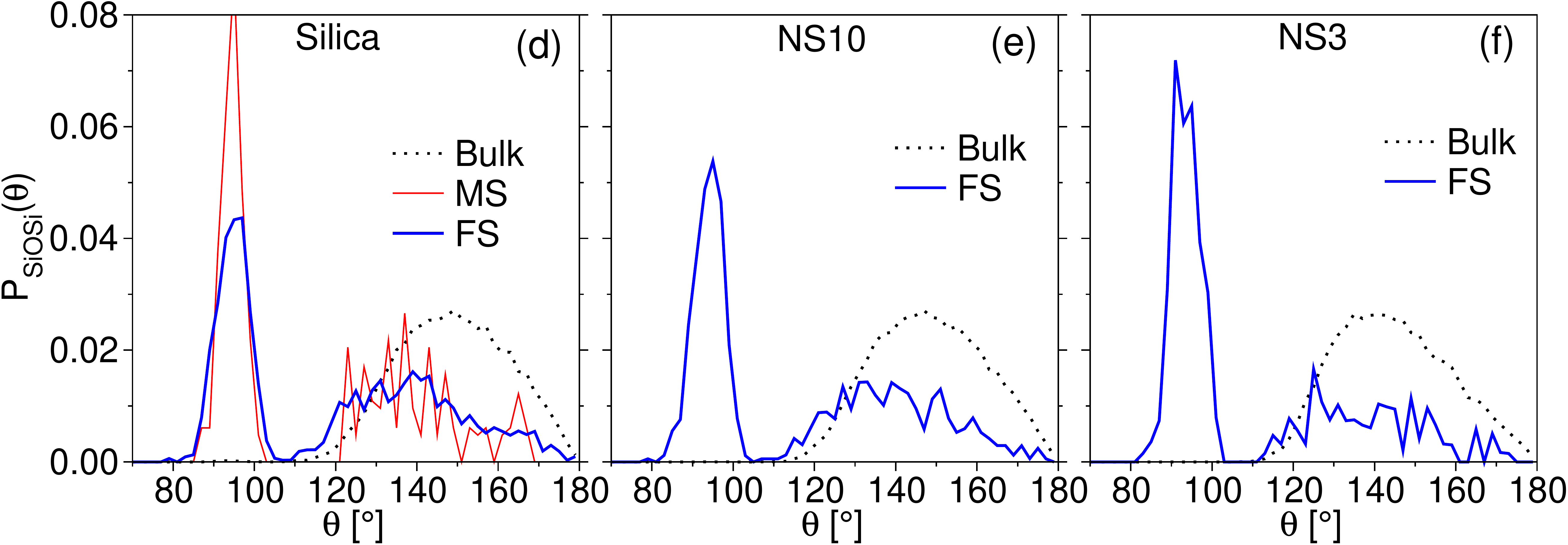}
\caption{Bond angle distribution. Upper and lower panels are for O-Si-O
and Si-O-Si angles, respectively. From left to right the compositions
are silica, NS10, and NS3, respectively.}
\label{fig:cl nsx-surf-BAD}
\end{figure*}

Figure~\ref{fig:cl nsx-surf-BAD} shows the bond angle distributions
for O-Si-O and Si-O-Si linkages. (Links are defined via the first
minimum in the corresponding radial distribution function.) For O-Si-O,
panels (a-c), one sees that the curves for the surfaces have a peak
at the intra-tetrahedral angle 109$\degree$ which is broader than the
one of the bulk. For the FS this peak is the widest, showing that the structure of this surface has the strongest disorder.  A further feature to be noticed is the
peak at $\approx85\degree$, notably for the FS curves. This peak is another
signature of the 2M-rings, in addition to the distance $r_{\rm
SiSi}=2.4$~\AA, Figs.~\ref{fig:cl nsx-surf-dist-PDF-si}(d-f). The presence
of these 2M-rings can also be seen in the distribution of the
(inter-tetrahedral) Si-O-Si angle, Figs.~\ref{fig:cl nsx-surf-BAD}(d-f),
in that one finds a marked peak at $\approx95\degree$. In fact this peak
is the dominant one for the FS, while it is completely absent in the bulk
sample, showing that the 2M-rings are very important structural motifs
for the FS. (Note that for NS10 and NS3 we do not show this distribution for the MS
since the quasi-absence of Si atoms makes that the distribution is
very noisy.) The abundance of the 2M-rings has also the consequence that the peak found
in the bulk system at around $150\degree$, stemming from rings of size
5-7,~\cite{sundararaman_new_2019} is shifted to smaller angles
and is reduced significantly in intensity.

Having discussed some of the structural properties of the surfaces for
three compositions we now focus on how the surface composition depends
on the Na concentration of the sample. From panel (a) of Fig.~\ref{fig:cl
nsx-surf-strcut-compo} one recognizes that for the MS the Na concentration is significantly higher than the value in the bulk (dashed
line), showing that the MS
is significantly enriched in sodium. Interestingly we find that the
Na$_2$O dependence of this concentration closely tracks the one of
the bulk (the lines are basically parallel). This means that once the
Na$_2$O concentration of the glass surpasses a certain amount (around
5\%) the surface becomes enriched in Na and increasing the Na$_2$O
concentration does not lead to a modification of the structure beyond
the trivially expected amount. For the FS we find a different behavior
in that increasing Na$_2$O gives rise to a (linear) increase in the
Na concentration but this time with a slope that is higher than the
one of the bulk. 
%This behavior is likely related to the fact that with increasing Na$_2$O the system becomes increasingly plastic and hence can relax in a substantial manner the defects that have been created by the fracture process, i.e.~the system will be abel to approach the composition of the MS. 
As a consequence, the Na fraction in the FS gradually approaches the one of the MS as the Na$_2$O concentration is increased. 
%As a consequence we find that for the FS at low Na$_2$O concentration the amount of Na at the surface is significantly smaller than the one found at the MS while at high Na$_2$O concentration there is basically no difference.
The same qualitative trends are observed for the concentration of oxygen:
Both type of surfaces have a O~concentration that is higher than the
one in the bulk but now the one for the MS is closer to the bulk curve
than the FS. 
%This indicates that in the MS the surface atoms do not only replace the Si atoms, but to some extent also the O atoms.

Also included in Fig.~\ref{fig:cl nsx-surf-strcut-compo}(a)
are the experimental data for the FS of NS3 as measured by
LEIS,~\cite{almeida_low-energy_2014} which shows the enrichment of
Na in the surface layer, in quantitative agreement with our results.
The experimental value for the O concentration is compatible with our
results, but due to the relatively large error bars of the experiment
data one cannot draw strong conclusions.

For the oxygen atoms one can distinguish between bridging (BO, bonded
to two Si) and nonbridging oxygen (NBO, bonded to only one Si) and their
fractions are depicted in Fig.~\ref{fig:cl nsx-surf-strcut-compo}(b). We
see that, for the bulk glasses, these concentrations show a linear dependence on the Na$_2$O
content and that for all compositions our data are in excellent
agreement with the results of XPS and nuclear magnetic resonance (NMR)
studies.~\cite{veal1982xps,maekawa1991structural,nesbitt_bridging_2011}
We observe that both type of surfaces are more abundant in NBO than
the bulk, and that this enrichment is more pronounced for the MS. The
concentration of NBO on the surface is directly related to the abundance
of Na: More Na on the surface results in the breaking of Si-O-Si linkage
thus creating more dangling Si-O$^-$ bonds, i.e. more NBO.

Figure~\ref{fig:cl nsx-surf-strcut-compo}(c) shows the fraction
of under-coordinated (3-fold) Si, Si$^3$, a typical structural
defects on glass surfaces. We note that the concentration of Si$^3$
is nearly zero for the MS. 
%, in agreement with the observation made in the context of Fig.~\ref{fig: nsx-surf-si-essi} that on this type of surface the concentration of Si is basically zero. 
In contrast to this,
the FS has a non-negligible amount of Si$^3$ and its concentration shows
an exponential dependence on Na$_2$O concentration. The presence of
Si$^3$ on the FS is due to the fact that the glass was fractured at room
temperature with a crack velocity on the order of 10$^3$~m/s, i.e.~dynamic
fracture.~\cite{zhang_thesis_2020,quinn2019terminal} Since at this $T$
the glass structure is practically frozen, the structural damage caused by the  fracture
can hardly be healed, leaving some Si atoms under-coordinated. The fact
that the fraction of Si$^3$ depends on the Na concentration demonstrates
the crucial role of Na in reducing and repairing the structural damages
during fracture.

%In the context of Fig.~\ref{fig:cl nsx-atom-num-frac-vapor} we have seen that the surface is slightly charged, an feature that will influence the behavior of atoms in the free space and which is also important for the reactivity of the surface. To quantify how this charge depends on the composition of the sample
Using the surface compositions, we have further calculated for the surfaces the per-atom atomic charge
$Q$, which is defined by

\begin{equation} 
Q=\sum f_\alpha q_\alpha \quad, 
\label{eq3}
\end{equation}

\noindent
where $f_\alpha$ and $q_\alpha$ are the fraction and charge
of atom species $\alpha$ ($\alpha \in {\rm O, Si, Na}$),
respectively. Figure~\ref{fig:cl nsx-surf-strcut-compo}(d) shows that
both the MS and FS are negatively charged, and that the negativity is more
pronounced for the MS than for the FS. With increasing Na$_2$O
concentration $|Q|$ diminishes significantly. These observations are
related to the fact that there are more oxygen ions on the surface
than is expected from the stochiometry, thus giving rise to a local
charge imbalance. With increasing Na concentration, sodium atoms will
propagate to the surface to compensate the negative charges, rationalizing
the decrease of $|Q|$ with increasing Na$_2$O concentration, although
even for the highest concentrations we consider the surface charge
remains significantly negative. The charge dependence on the sodium
content is stronger for the MS than for the FS since the former is better
equilibrated and hence the sodium atoms have a higher probability to
reach this surface, in agreement with the higher Na fraction found in the MS (Fig.~\ref{fig:cl nsx-surf-strcut-compo}a). %For Na$_2$O concentration above 12\% the surface charge for the FS is very close to the one of the MS. since the Na atoms do not have to move far to reach the surface and hence can do so even for the FS.

\subsection{Depth profiles}  
\label{sec: depth-profiles}

Having defined the outermost surface layer we can now
investigate how the composition of the sample changes as a function
of the depth $r$, i.e.~over what distance the presence of the surface
affects the properties of the glass sample. Note that in the following we
define this distance $r$ as the length of the shortest path from a given
atom to any atom on the surface and thus $r=0$ represents the surface
monolayer.  

\begin{figure*}[htp]
\centering
\includegraphics[width=0.85\textwidth]{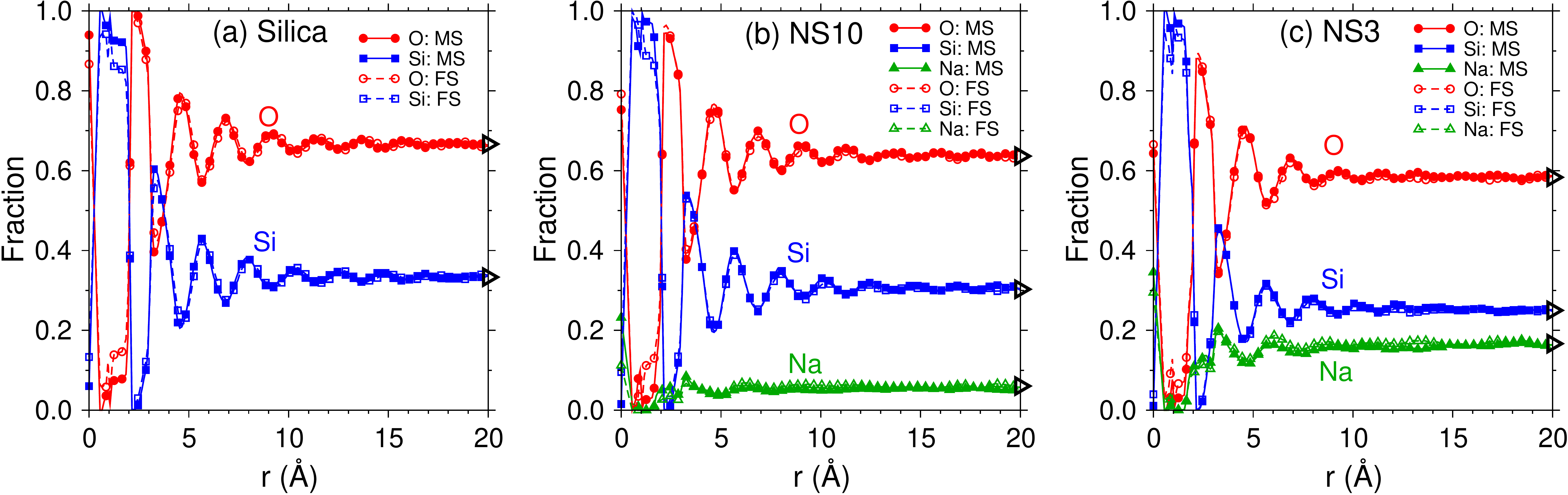}
\includegraphics[width=0.85\textwidth]{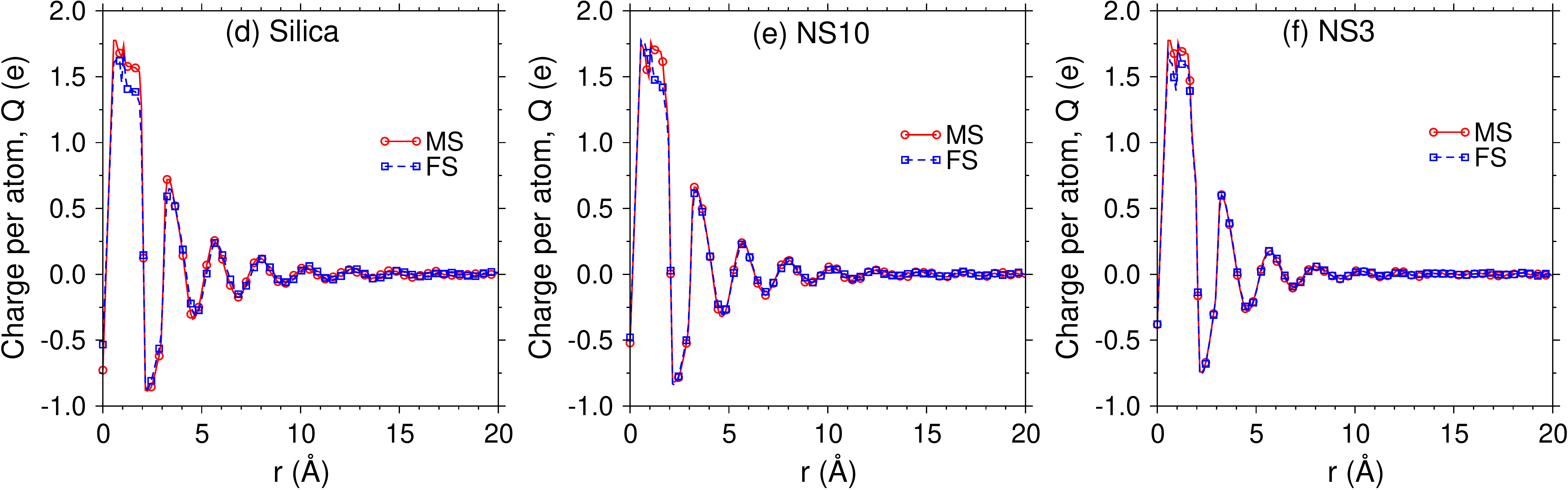}
\includegraphics[width=0.85\textwidth]{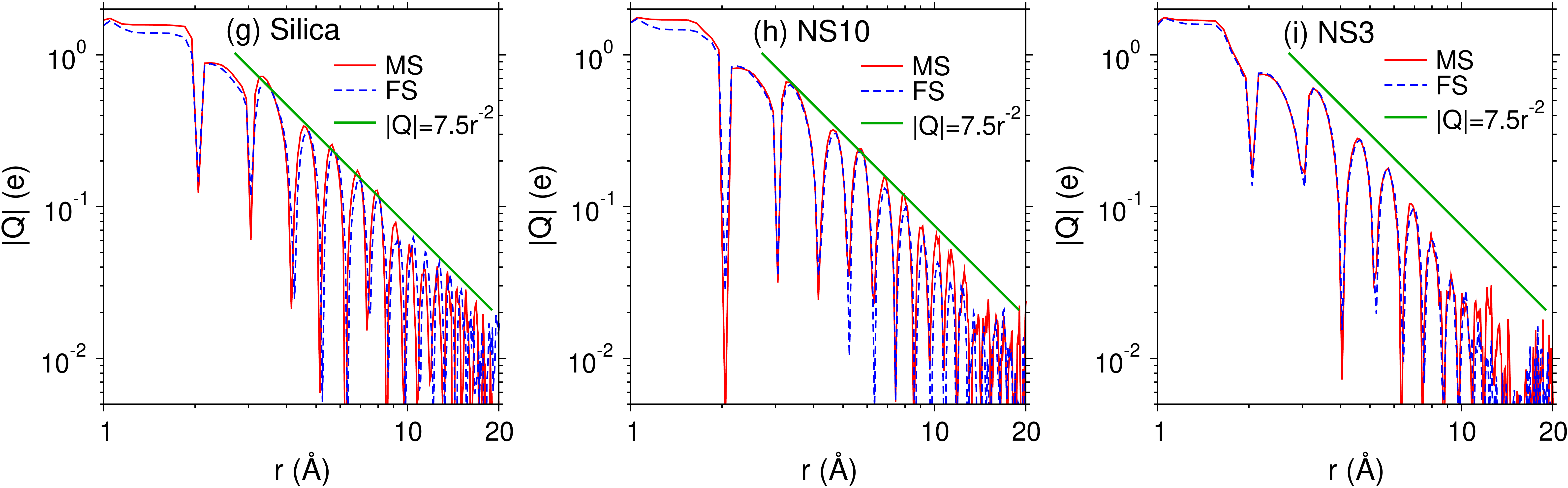}
\includegraphics[width=0.85\textwidth]{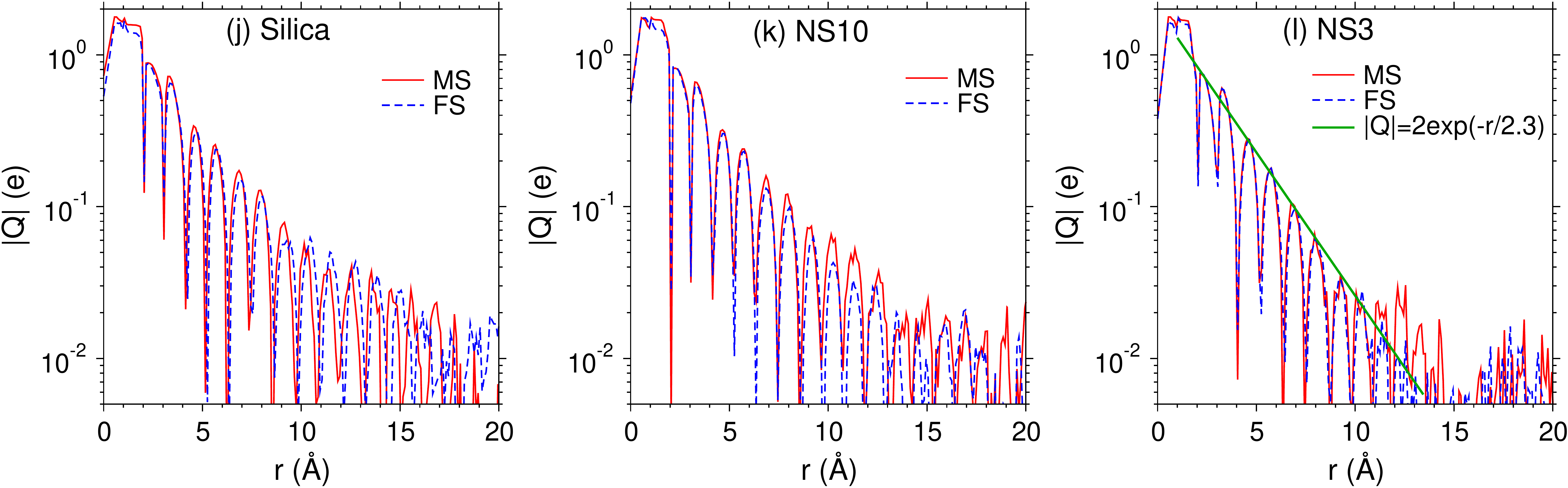}
\caption{(a-c): Depth profiles of elemental concentrations
with respect to the monolayer surfaces for silica, NS10, and NS3, respectively. In
practice, the composition at distance $r$ is the mean of a 1.1~\AA\
thick layer. (d-f): Depth profiles of the per-atom atomic charge. (g-i):
Log-log plots of the data shown in panels (d-f). Note that we now show
the absolute value of the atomic charge.  The green solid lines in the
graphs are guides to the eye and have the slope $-2$. % {\color{blue}WK: I hope that a slope of -2 is compatible with the data. Else we have tochange the numbers.} 
(j-l): Log-linear plots of the data shown in panels (d-f). The green solid line in (l) indicate the exponential decay of $|Q|$ with distance $r$. }
\label{fig:cl nsx-surf-depth-profile}
\end{figure*}

Figures~\ref{fig:cl nsx-surf-depth-profile}(a-c) show the
concentration profiles of various atomic species as a function of $r$ for
the two types of surfaces. These graphs demonstrate that for these three
compositions the curves for the FS and MS are very similar if $r$ exceeds
2-3~\AA, while noticable differences are seen at smaller distances. 
With increasing $r$, the oxygen concentration drops very quickly if $r$ reaches 1~\AA~while the concentration of
Si increases strongly. This signals that the atomic layer right below
the surface is dominated by silicon atoms, in agreement with previous
studies.~\cite{roder_structure_2001,rarivomanantsoa_classical_2001,halbert2018modelling}
%By comparing the elemental concentrations for the MS with the one from the FS one sees that for distances smaller than around 1~\AA~the peaks for the MS are more pronounced, in agreement with the arguments put forwardabove that the MS is more relaxed than the FS. 
The differences between the curves for the MS and FS become
invisible for distances $r>2$~\AA, i.e.~beyond this distance
the density profiles do not depend on how the surface
has been created. %Although this result does not imply that beyond that distance there are no structural differences between the two type of samples, it indicates that the difference is not very pronounced. 
If $r$ is increased further, the elemental concentrations approach the value in the bulk (marked
by triangles on the right ordinate) and for distances around
20~\AA~the curves reach the bulk values within 1\% deviation. Note that
for the systems containing sodium, panels (b) and (c), the decay of the Na profiles is faster
than the one for Si and O, a result that is related to the high mobility
of the Na atoms which allows this species to screen quickly the perturbations generated by the surface.

Figures~\ref{fig:cl nsx-surf-depth-profile}(d-f) show the $r-$dependence
of the per-atom atomic charge, defined in Eq.~(\ref{eq3}), for the two
surfaces. As $r$ increases, one finds alternating peaks/valleys,
a result that is directly related to the variation of atomic fractions
shown in panels (a-c). In addition, we notice that the charge fluctuations
seem to decay faster if the Na content in the glass is increased. To
understand better the $r-$dependence of $Q$, we have replotted the data
on log-log (Figs.~\ref{fig:cl nsx-surf-depth-profile}(g-i)) and semi-log
(Figs.~\ref{fig:cl nsx-surf-depth-profile}(j-l)) scales. (Note that
now we plot the absolute value of $Q$.) These graphs allow to recognize
that, within the noise of the data, the charge at intermediate and large
distances is independent of the type of surface, despite the fact that the
charge on the first layer is more negative for the MS than for the FS,
see Fig.~\ref{fig:cl nsx-surf-strcut-compo}(d). For the case of silica,
panel (g), we find that the decay is nicely descibed by a power-law with an
exponent -2. If one adds a bit of sodium, NS10 in panel (h), the signal
at intermediate range is compatible with the same power-law, but for
distances larger than 8~\AA~one spots deviations. For systems with high
sodium content, NS3 in panel (i), the power-law is no longer a good description for the decay.

The log-linear plot in panel (j) confirms that for silica the decay of $|Q|$
is slower than an exponential. With the addition of Na, NS10 in panel (k),
the decay becomes faster but within the limited $r-$range of the data it
is not possible to identify the functional form of the decay. However, for
the case of NS3, panel (l), things become clearer in that the signal can be
nicely described by a straight line, i.e.~the decay is an exponential with
a decay length $\approx2.3$~\AA.  Hence we can conclude from these graphs
that the addition of sodium transforms the power-law decay observed in
silica into an exponential decrease, i.e.~the network modifier is able
to restore the charge balance already at short distances (about 10~\AA).

\subsection{Vibrational properties}

The quantities discussed so far to characterize the surface are closely related to its structure.  In real experiments it is, however, not easy to access this type of information since, e.g., scattering techniques are hampered by a lack of scattering volume or spatial resolution. %~\cite{XXX}. ZZ: not clear which reference Walter implies.
As a consequence one often relies on spectroscopic techniques to investigate surface properties since such measurements allow to pick up a signal even if the probe volume is small. 
In order to make a connection of the structural
properties with the spectroscopic properties of the
samples, we discuss in the present section its vibrational features.

Although simulations using classical potentials have often difficulties to
give a reliable description of the vibrational spectra,~\cite{benoit_vibrational_2002}
the potential used in the present work has been found to be able to
reproduce well the vibrational features of soda-silicate glasses in the
bulk.~\cite{sundararaman_new_2018,sundararaman_new_2019} Therefore it
can be expected that this potential gives also reasonable values for the vibrational
frequencies of the atoms close to the surface.

To calculate the vibrational density of states (VDOS), $g(\omega)$, of
the system we have quenched the glasses to 5~K. Since the samples were
already at 300~K, i.e.~well below their glass transition temperature,
this quench can be done with a high cooling rate without affecting
the $g(\omega)$.  At this low temperature the motion of the atoms can
be considered to be harmonic and hence the VDOS can be obtained by
calculating the time Fourier transform of the velocity-autocorrelation
function:~\cite{dove1993introduction}

\begin{equation} 
g(\omega)=\frac{1}{Nk_{\rm B}T}\sum_j
m_j\int_{-\infty}^{\infty}dt \ {\rm exp}(i\omega t)\big \langle
\*v_j(t)\*v_j(0) \big \rangle, 
\label{eq4}
\end{equation} 

\noindent
where $\omega$ is the frequency, $N$ is the number of atoms, $k_{\rm B}$
is the Boltzmann constant, and $m_j$ and $\*v_j(t)$ are respectively
the mass and velocity of atom $j$ at time $t$. This way to calculate
$g(\omega)$ has the advantage that it is computationally inexpensive
and also allows the decomposition of the VDOS into various species
(atoms, groups of atoms, etc.) by considering on the right hand side of
Eq.~(\ref{eq4}) the corresponding terms.

\begin{figure}[ht]
\centering
\includegraphics[width=0.47\textwidth]{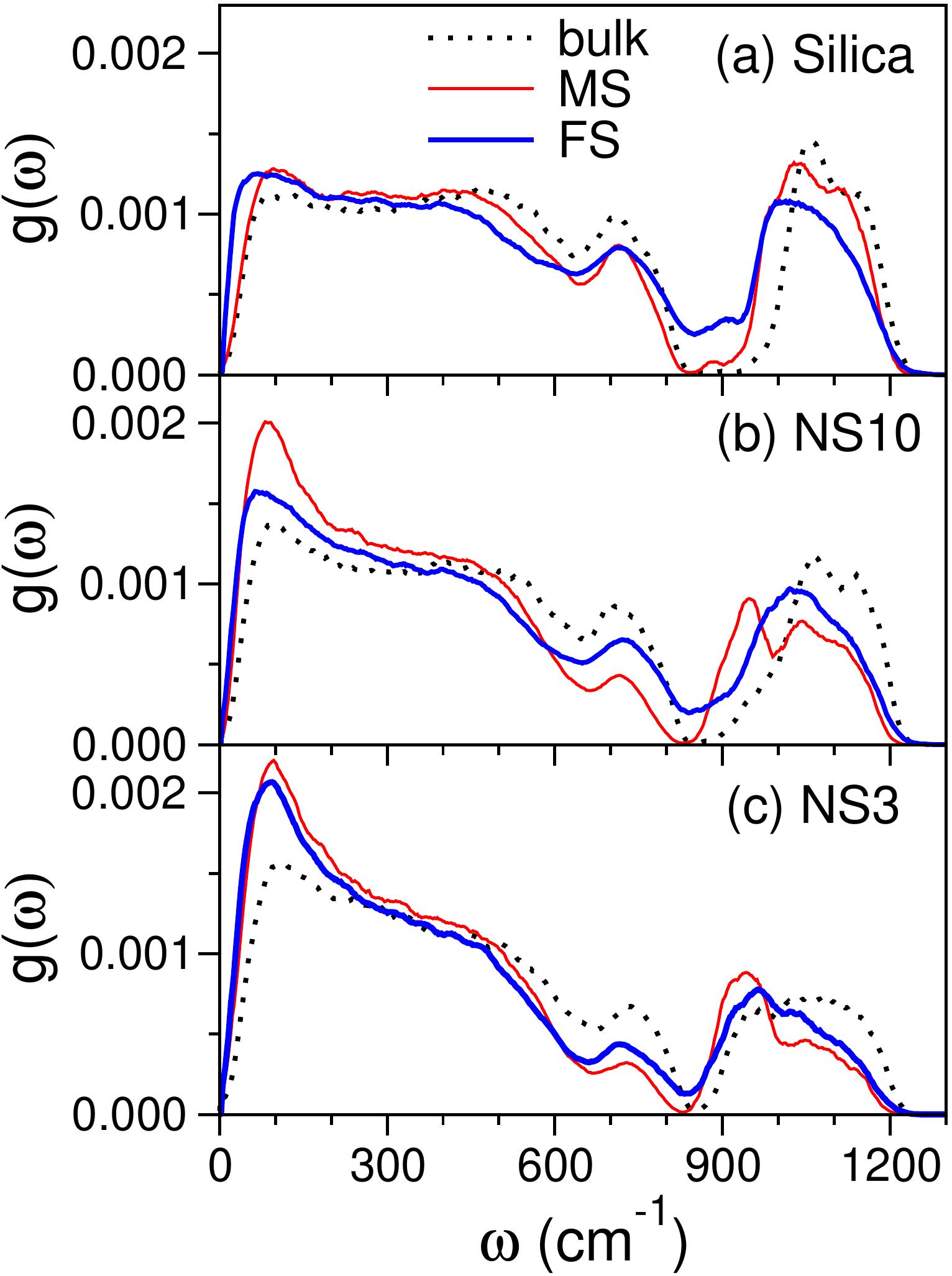}
\caption{Total vibrational density of states at 5~K. (a-c)
are for silica, NS10 and NS3 glasses, respectively. All VDOS curves are
normalized to unity. %The bulk data correspond to a cube (with side length $\approx6$~nm) in the geometrical center of the sandwich glasses.
 }
\label{fig:cl nsx-vdos-tot}
\end{figure}

Figure~\ref{fig:cl nsx-vdos-tot} shows the frequency dependence of the
VDOS, normalized to unity, for three glass compositions. 
%and that the spectra for the FS and MS will depend on the fraction of atoms that are close to the surface, i.e.~on the geometry of the sample, while the curve labeled ``bulk'' can be expected to be independent of this geometry. Despite this dependency it is possible to recognize the effect of the surface on the VDOS and hence the presented $g(\omega)$ is a useful quantity.
For the case of bulk SiO$_2$, dotted line in panel (a), it has been
documented that  the broad band with $\omega \leq 800$~cm$^{-1}$ is
due to bending and rocking modes of O with respect to Si, whereas the
high-frequency band with $\omega \geq 950$~cm$^{-1}$ originates from
the (symmetric/asymmetric) stretching vibrations within the [SiO$_4$]
units.~\cite{wilson1996polarization,sarnthein1997origin,taraskin_nature_1997}
For the MS one finds that the high frequency band has the same shape as
the one of the bulk, but it is shifted to somewhat lower frequencies,
showing that the presence of the surface makes the intra-tetrahedral
vibrations a bit softer, likely due to the fact that there are less
(effective) constraints on the atoms, but also because of the presence of NBO (see
below).  The bands at around 500~cm$^{-1}$ and 720~cm$^{-1}$ have a lower
intensity than the ones in the bulk and below we will see that this is
due to the increased number of NBO.~\cite{zotov_calculation_1999} This
enhanced concentration of NBO is also the reason for the increase
of the peak at around 100~cm$^{-1}$ since in this frequency range
one has significant contributions from the rotation of [SiO$_4$]
units.~\cite{taraskin_nature_1997} 
For the FS we see that the gap in $g(\omega)$ at around 850~cm$^{-1}$
starts to fill up since, see below, in this range of $\omega$ the edge
sharing BO have a significant contribution to $g(\omega)$. The band at
high frequencies has less structure than the one for the bulk or the MS,
a result that is related to the increased disorder in the tetrahedral units. The rest of
the spectrum is qualitatively similar to the one for the MS, except that
the peak at around 100~cm$^{-1}$ is shifted to even lower frequencies
since the structural units at the surface have a decreased connectivity.

For the bulk glasses containing Na, Figs.~\ref{fig:cl nsx-vdos-tot}(b) and
(c), we note firstly that the intensities of the main bands at intermediate
and high frequencies are reduced with respect to the ones in silica. In
contrast to this, one observes the presence of a pronounced peak at
$\omega<200$~cm$^{-1}$ the intensity of which increases with increasing
Na concentration. This peak is related to the vibrational modes of the Na atoms.~\cite{kilymis_vibrational_2019} For the surfaces this low-frequency peak is significantly higher than the one for the bulk, in agreement with the fact discussed above that the surfaces have higher concentration
of Na and NBO~(Fig.~\ref{fig:cl nsx-surf-strcut-compo}).
The increase of Na content also leads to a substantial reduction
of the peak at around $720$~cm$^{-1}$, since the presence of Na
reduces the number of corner-sharing bridging oxygen atoms (see
below). For the high frequency band we recognize that the presence
of a surface leads to a significant change in the shape of the
peaks, excitations that are related to complex Si-O and Si-NBO
motions.~\cite{zotov_calculation_1999,kilymis_vibrational_2019}

\begin{figure*}[!t]
\centering
\includegraphics[width=0.95\textwidth]{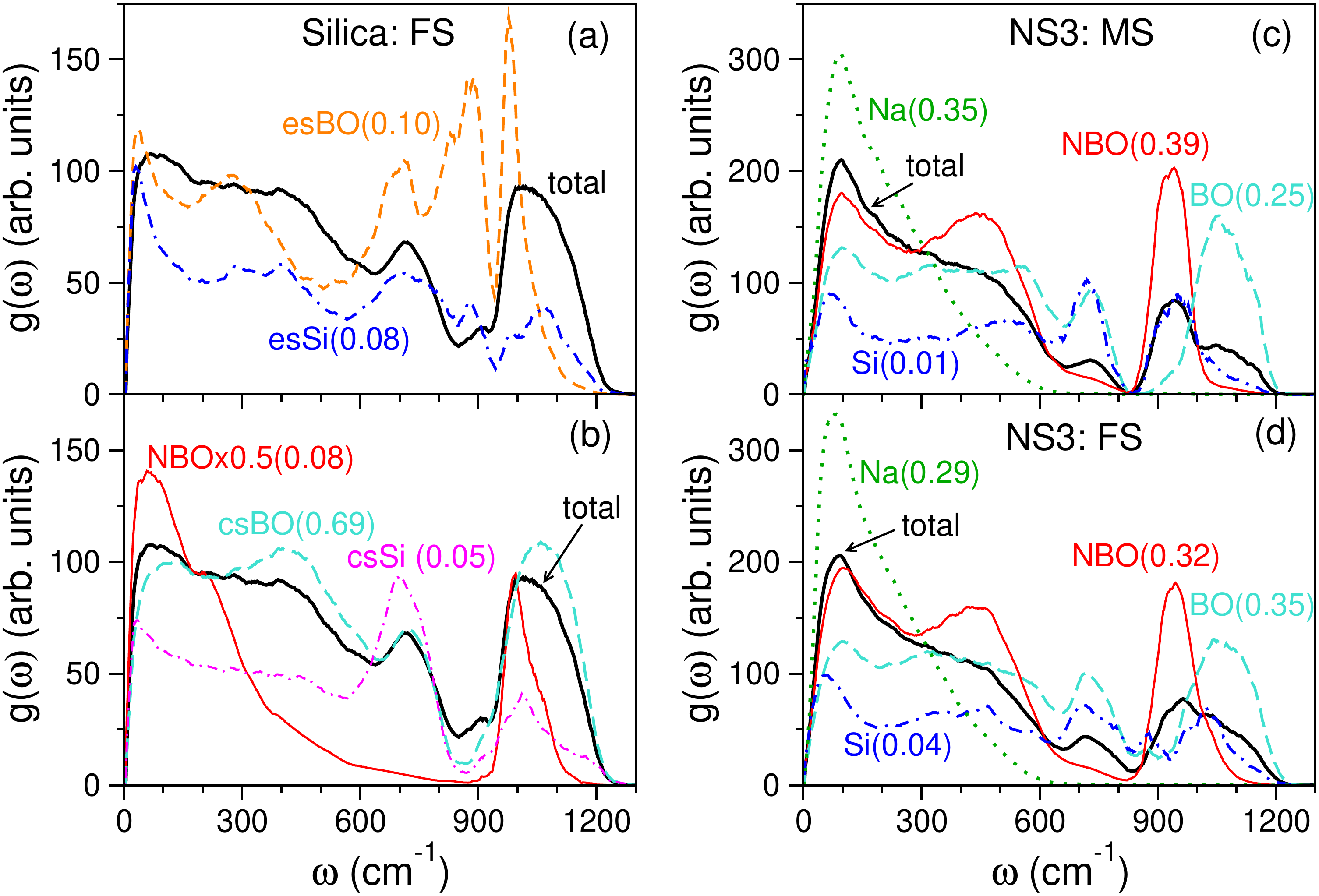}
\caption{Per-atom VDOS of various atomic species. 
%{\color{blue}WK: I think that it would be better to normalize the distributions to 1.0 so that it is clear than we can compare the curves in the different panel.} %{\color{red}ZZ: I think it's more fair to compare the curves in the present form since there are per-atom quantity. Normalizing the partials to unity will wash out their real vibrations at different frequencies.}
The number in the parenthesis is the fraction of a given species with
respect to the total. (a) and (b) are for the FS of silica. The curve for
NBO is multiplied by 0.5 to allow better comparison. %{\color{blue}{WK:Remove the star before NBO and instead write ``NBO$\times 0.5$''}}
(c) and (d) are for the MS and FS of NS3, respectively.}
\label{fig:cl nsx-vdos-partials}
\end{figure*}

In order to understand in a more quantitative manner the reason for these
modifications it is useful to decompose the VDOS into the contributions
of the various structural elements.
Figures~\ref{fig:cl nsx-vdos-partials}(a) and (b) present the various
partial VDOS for the FS of the silica glass. The structural elements we
consider are the edge-sharing and corner-sharing BO (esBO and csBO,
respectively), the edge-sharing and corner-sharing silicon atoms
(esSi and csSi, respectively) as well as the NBO.  (For the sake
of clarity, these five partials are shown separately in panels (a)
and (b).) Also included in the graphs are the total VDOS (the same as
shown in Fig.~\ref{fig:cl nsx-vdos-tot} but here it is normalized with respect to the number of atoms). The total VDOS is the weighted sum of
the partials, with weights that are stated next to each curve. From
these distributions one recognizes that around 950~cm$^{-1}$
the NBO as well as the esBO have a sharp peak. This rationalizes thus
the observation that in Fig.~\ref{fig:cl nsx-vdos-tot}(a) the high frequency
band of the FS is shifted to lower frequency with respect to the bulk system.
The disappearance of the gap at around 850~cm$^{-1}$ can now be explained
by a signal in the VDOS of the esBO since these 2M-rings have a marked
peak at $\approx880$~cm$^{-1}$ (panel a). We mention that experimental
studies have reported that the presence of 2M-rings on the surface
of $\beta$-cristobalite generate two strong infrared bands at 888 and
908~cm$^{-1}$.~\cite{morrow1976infrared,bunker1989infrared,grabbe1995strained}
The vibrational fingerprint of 2M-ring found in our simulations
is thus in excellent agreement with these experimental findings. To
the best of our our knowledge this is the first time the vibrational
signature of the 2M-ring structure is correctly predicted from classical
simulations. The fact that we did not observe a double peak might be
attributed to the disordered structure of glass. (The vibrational spectra
of glass are generally broader than their crystalline counterparts. Since
the two characteristic peaks for 2M-rings are very close to each other,
it is likely that they merge together to form a broader band in the case of glass.)

Furthermore, we note that the 2M-rings affect also the total spectrum at
lower frequencies. The spectrum for the csBO have a peak at around
420~cm$^{-1}$ (Fig.~\ref{fig:cl nsx-vdos-tot}b) while for the esBO this peak is shifted to around
300~cm$^{-1}$, which rationalizes the different shape of the total VDOS
in this frequency range (see Fig.~\ref{fig:cl nsx-vdos-tot}a). 
%{\color{blue}{WK: Before we had the following text, but I don't agree with it: ``..., the vibrations of the corner-sharing atoms are also affected, particularly for the ones in connection with the 2M rings. This explains the result that $g(\omega)$ is larger than 0 at $\approx900$~cm$^{-1}$ for csSi and csBO, Fig.~\ref{fig:cl nsx-vdos-partials}(b).''}}
Panel (b) also shows that the NBO have a marked peak at around
80~cm$^{-1}$. This feature explains thus why the total VDOS for the
FS has an enhanced intensity at low frequencies with respect to the
bulk sample.

We show in Figs.~\ref{fig:cl nsx-vdos-partials}(c) and (d) the
partial VDOS for the MS and FS of the Na-containing NS3 glass. For both
systems we find that the Na VDOS has a pronounced peak at small frequencies,
in qualitative agreement with previous simulation results of bulk
systems.~\cite{kilymis_vibrational_2019} Note that for the FS this
peak is somewhat higher than the one for the MS. This is likely due to
the fact that for the MS the local Na concentration is higher than for
the FS, thus allowing environments of the Na atoms that are more diverse
and hence a peak that is broader.
Furthermore we see that the curves for the NBO have a pronounced peak at
$\approx950$~cm$^{-1}$ while the one for BO have a strong contibution
at $\omega>1000$~cm$^{-1}$. These two peaks explain thus the observed
double peak structure in NS3 in the high frequency band of the NS3 system
having a surface, see Fig.~\ref{fig:cl nsx-vdos-tot}(c). %(Note that for the NS3 bulk system the height of these two peaks are almost the same and hence one does not see a double peak structure but only a broad band.)

Finally we note that for the FS we can also detect a small peak at $\omega\approx860$~cm$^{-1}$ in
the VDOS for the Si and BO, panel (d), while this feature is absent in
the spectra for the MS. These peaks are related to the 2M-rings of the
surface which are present on the FS but not on the MS (see Fig.~\ref{fig:
nsx-surf-si-essi}). So this peak is able to tell whether a surface has
been generated from a quench of the melt or the fracture of a glass
sample.

% \begin{figure*}[ht]
% \centering
% \includegraphics[width=\textwidth]{vdos-ns3-partials-peratom-compare.eps}
% \caption{Per-atom VDOS of various atomic species. This figure will not
% go to the final paper and it shows here only for us.}
% \label{fig:cl nsx-peratom}
% \end{figure*}

\clearpage
\section{Summary and Conclusions}
\label{sec_IV}

In this work we have probed how the properties of glass surfaces depend
on the composition of the system and the type of surface considered
(melt-formed surface or fracture surface). By analyzing the surface monolayer we have found that, independent of the Na$_2$O concentration, there
are significant differences between the MS and FS; the latter have quite a few structural defects such as two-membered rings and under-coordinated Si while the former have none, except for the case of silica. 
This shows that, for the MS, the annealing of the structure as made possible by the high-temperature equilibration and 
slow cooling allows to avoid these energetically unfavorable
structures, {\it if} Na are present. Also the composition of the
first atomic layer depends on the type of sample considered in that the
elemental concentrations of the FS show almost a linear dependence on
Na$_2$O content in the glass, whereas the ones for the MS behave very differently. Both the MS and FS are negatively charged. For the glasses with low Na$_2$O concentration, this charge is relatively large but its negativity decreases with increasing Na$_2$O
content. This effect can be expected to be important for the chemical
reactivity of the surface since local Na fluctuations will result in local
fluctuations in the charge.  

Since the presence of the surface creates
a gradient in the composition we have probed how these fluctuations
depend on the type of surface and the glass composition. Surprisingly, we find
that beyond the second atomic layer below the surface, i.e.~$r>2$~\AA, there is no noticeable
difference between the compositional fluctuations measured in the FS and MS systems. We emphasize, however, that this result
does not mean that the structure of the glass farther than,
say, a few \AA~from the surface is independent of the way the surface
was generated. Interestingly, the functional form for the decay of the elemental concentration depends on the Na$_2$O concentration in the
sample: For low Na$_2$O concentration it is a power-law while for high concentration one finds an exponential decay. The slower (power-law) decay indicates the
high local frustration of the system which permits only a gradual
healing of the structural perturbation due to the presence of free surface.

The vibrational density of states show that the systems with free surfaces
have on average atomic vibrations at lower frequencies than the one found
in the bulk glass. By analyzing the partial VDOS we show that this
softening of the atomic vibrations is due to the increased number of NBO close to the surface and
also to the higher concentration of Na atoms. Interestingly we find in
the spectrum of the FS a weak but clear signal of the two-membered rings at the frequency of $\approx880$~cm$^{-1}$,
while no such signal exists for the MS. This result therefore permits
to use spectroscopic methods to decide whether or not a given surface
has been generated from a fracture process.

\section{Acknowledgements}
We thank H. Jain for discussions. Z.Z. acknowledges financial support by China Scholarship Council (NO. 201606050112).
W.K. is member of the Institut Universitaire de France.
This work was granted access to the HPC resources
of CINES under the allocation A0050907572
attributed by GENCI (Grand Equipement National de Calcul
Intensif).
\medskip

\noindent {\bf Data Availability Statement:} The data that support the findings of this study are available from the corresponding author upon reasonable request. 

\normalem  %use the original definition of \emph for the list of references, intstead of a underline
%\bibliography{jcp}

\end{document}